\documentclass[journal=eds]{CUP-JNL-DTM}%

\usepackage{graphicx}
\usepackage{multicol,multirow}
\usepackage{amsmath,amssymb,amsfonts}
\usepackage{mathrsfs}
\usepackage{amsthm}
\usepackage{rotating}
\usepackage{appendix}
\usepackage{ifpdf}
\usepackage[T1]{fontenc}
\usepackage{newtxtext}
\usepackage{newtxmath}
\usepackage{textcomp}
\usepackage{xcolor}
\usepackage{lipsum}
\usepackage[colorlinks,allcolors=blue]{hyperref}
\usepackage{pdflscape}
\usepackage{lineno}
\usepackage{graphbox}
\usepackage[section,above,below]{placeins}

\theoremstyle{definition}

\numberwithin{equation}{section}

\usepackage[dvipsnames]{xcolor}

\jname{Environmental Data Science}
\articletype{APPLICATION PAPER}
\jyear{2026}

\begin{document}
\begin{Frontmatter}

\title[Article Title]{Calibrated Conformal Prediction Intervals for Microphysical Process Rates}

\author[1]{Miriam Simm}
\author[1]{Corinna Hoose}
\author[2,3]{Tom Beucler}

\address[1]{\orgdiv{Institute of Meteorology and Climate Research Troposphere Research}, \orgname{Karlsruhe Institute of Technology}, \orgaddress{\city{Karlsruhe},  \country{Germany}}. \email{miriam.simm@kit.edu}}
\address[2]{\orgdiv{Faculty of Geosciences and Environment}, \orgname{University of Lausanne}, \orgaddress{\city{Lausanne}, \country{Switzerland}}}
\address[3]{\orgdiv{Expertise Center for Climate Extremes}, \orgname{University of Lausanne}, \orgaddress{\city{Lausanne}, \country{Switzerland}}}

\authormark{Miriam Simm et al.}


\abstract{
Conformal prediction can yield statistically valid prediction intervals for any regression model, with no model modifications and small computational costs. To assess its practical value, we apply conformal methods to quantify uncertainty in machine learning emulators of six microphysical process rates. Microphysical process rates describe small-scale processes in atmospheric clouds such as precipitation formation and aerosol-cloud interactions, and help understand weather and climate.
The emulators are trained on simulation output from the ICOsahedral Nonhydrostatic (ICON) model in a limited-area numerical weather prediction configuration. We compare split conformal prediction for deterministic emulators with conformalized quantile regression for quantile regression emulators. Both conformal prediction methods yield well-calibrated and sharp prediction intervals on average, but conformalized quantile regression provides more consistent intervals across several orders of magnitude, making it preferable for the uncertainty quantification of climate variables.
}
\end{Frontmatter}

\section*{Impact Statement}
Robust uncertainty quantification is important for the reliability of model predictions. Conformal prediction is a model-agnostic and distribution-free framework to obtain statistically valid prediction intervals, but remains underused in climate science. By applying and evaluating conformal prediction methods for microphysical process rates spanning several orders of magnitude, we demonstrate their potential for inexpensive uncertainty quantification in weather and climate modeling.

\section{Introduction}
In atmospheric prediction, purely deterministic methods increasingly reveal their limitations. For subgrid-scale parameterizations, grid-scale state variables may not uniquely determine the net effect of unresolved processes \citep{palmerStochasticWeatherClimate2019,christensenMachineLearningStochastic2024}. This motivates probabilistic frameworks, stochastic parameterizations, and robust uncertainty quantification \citep{christensenMachineLearningStochastic2024, haynesCreatingEvaluatingUncertainty2023, schreckEvidentialDeepLearning2024, mansfield2025epistemicaleatoricuncertaintyquantification}, often implemented through ensemble approaches \citep[e.g.,][]{behrensSimulatingAtmosphericProcesses2025, shinParameterizationStochasticallyEntraining2022, gagneMachineLearningStochastic2020} that can be computationally costly.

As a computationally inexpensive alternative, conformal prediction (CP) \citep{vovkLeaningByTransduction1998,vovkAlgorithmicLearningRandom2022,angelopoulosGentleIntroductionConformal2022} provides distribution-free prediction intervals with finite-sample validity for any regression model, under the assumption that the data is exchangeable, which is weaker than the independent and identically distributed assumption commonly used in machine learning (ML). Yet, CP has seen limited use in weather and climate applications \citep[e.g.,][]{gopakumarUncertaintyQuantificationSurrogate2025,mortierValidPredictionIntervals2025}, in part because spatio-temporal dependence in geophysical data can violate exchangeability. Extending conformal methods to such settings remains an active research area \citep[e.g.,][]{xuConformalPredictionInterval2021, gibbs2021adaptiveconformalinferencedistribution, sun2022conformalmethodsquantifyinguncertainty}.

In numerical models, the spatial and temporal discretization separates resolved from unresolved scales. Subgrid-scale physical processes must therefore be represented through parameterizations, which are a major source of model uncertainty \citep{boucherClimateChange2014}. Cloud microphysics describes phase transitions of condensed water in the atmosphere and interactions among liquid and frozen particles, water vapor, and aerosols. These processes control precipitation formation and intensity, impact storm evolution, and, through latent heating and cooling, influence cloud dynamics and large-scale cloud structure. Microphysical properties also affect radiative transfer and thus the Earth's radiation budget \citep{gettemlmannCloudMicrophysicsScales2019, morrisonConfrontingChallengeModeling2020, lambPerspectivesSystematicCloud2026}. Accurate representation of cloud microphysics remains particularly challenging due to its inherent complexity and nonlinearity, as well as incomplete process-level understanding \citep{zelinkaCausesHigherClimate2020,khainRepresentationMicrophysicalProcesses2015}. Improved understanding of cloud processes will help to reduce persistent uncertainties and advance the development of ML-based models \citep{lambPerspectivesSystematicCloud2026}. Microphysical process rates (MPRs), computed internally to update the prognostic cloud variables, would provide detailed insights into cloud processes and support model development. Yet, storage constraints typically preclude saving MPR output from large-scale high-resolution simulations, because MPRs comprise a two-digit number of three-dimensional variables. Moreover, offline recalculations of process rates are generally not accurate because of a temporal mismatch between the prognostic cloud variables in the model output and the MPRs computed during time stepping.

Here, we train machine learning models to emulate the computation of six MPRs from a two-moment bulk microphysics scheme \citep{seifertbehengTwomomentCloudMicrophysics2006} and quantify predictive uncertainty with conformal prediction methods. Using ML emulators forced by atmospheric state variables from limited-area ICOsahedral Nonhydrostatic (ICON) simulations \citep{zanglICONICOsahedralNonhydrostatic2015}, we (i) apply split conformal prediction to calibrate prediction intervals for deterministic emulators, (ii) train quantile regression emulators and calibrate their intervals via conformalized quantile regression \citep{romanoConformalizedQuantileRegression2019}, and (iii) benchmark interval calibration and sharpness across emulators and conformal prediction methods. 

The structure is as follows. Section~\ref{sec:conformal_prediction} reviews conformal prediction and conformalized quantile regression. Section~\ref{sec:methods} describes the microphysical process rates, the ICON model output-based datasets, and the training and calibration procedure. Section~\ref{sec:results} presents the results, and Section~\ref{sec:conclusions} concludes. 

\section{Theory: Conformal prediction}\label{sec:conformal_prediction} 
\begin{figure}
    \centering
    \includegraphics[width=\textwidth]{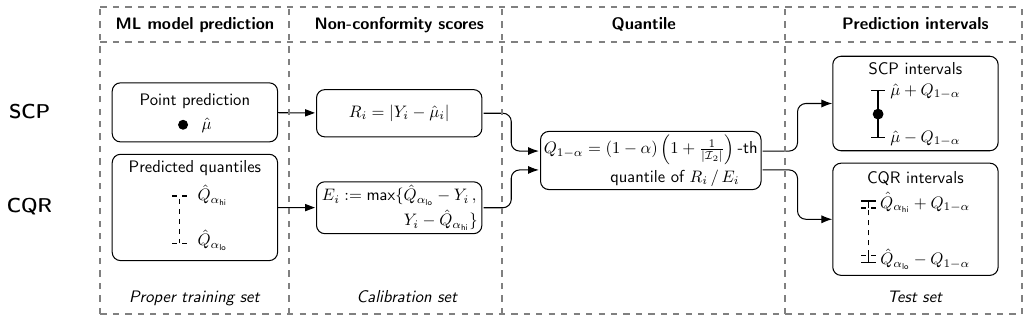}
    \caption{Conformal prediction framework: split conformal prediction (top row) and conformalized quantile regression (bottom row)}
    \label{fig:cp_algorithm}
\end{figure}

Conformal prediction (CP) provides distribution-free prediction intervals for any regression or classification model, which contain the true value of the target with a pre-defined probability, requiring only data exchangeability \citep{vovkAlgorithmicLearningRandom2022}. While originally formulated in a transductive setting \citep{vovkLeaningByTransduction1998}, ``full'' CP is computationally expensive because it requires retraining the underlying model for each test point. We hence focus on inductive (``split'') CP, which is computationally efficient but requires splitting the data into a proper training set and a calibration set \citep{leiDistributionFreePredictiveInference2018}. In split CP (Figure~\ref{fig:cp_algorithm}), a point-prediction model is trained on the proper training set, and non-conformity scores, which measure the prediction error, are computed on the calibration set. The $(1-\alpha)$-quantile of these scores sets the width of the prediction intervals. Here, $\alpha$ is a user-specified miscoverage rate, which quantifies the probability that the true value falls outside the prediction interval. This yields prediction intervals that, by construction, marginally contain the true target value with $100(1-\alpha)\%$ probability. Although this guarantee holds independently of model skill, data distribution and non-conformity score function, interval informativeness mainly depends on the choice of score function \citep{angelopoulosGentleIntroductionConformal2022}. 

\subsection{Split conformal prediction without retraining (SCP)}
More formally, given $n$ training samples $\{(X_i, Y_i)\}_{i=1}^n$, the main objective is to construct the prediction interval $\mathcal{C}(X_{n+1}) \subseteq \mathbb{R}$ for a test point $X_{n+1}$. The interval $\mathcal{C}$ contains the unknown value $Y_{n+1}$ with the miscoverage rate $\alpha \in (0,1)$, satisfying the marginal coverage guarantee
\begin{equation}\label{eq:coverage_guarantee}
    \mathbb{P}\left\{Y_{n+1} \in \mathcal{C}(X_{n+1})\right\} \geq 1-\alpha \, ,
\end{equation}
for any joint distribution of the feature vectors $X \in \mathbb{R}^p$ of dimension $p \geq 1$ and labels $Y \in \mathbb{R}$ and any sample size $n$ with the assumption that all samples $\{(X_i, Y_i)\}_{i=1}^{n+1}$ are drawn exchangeably from the joint distribution.

The procedure for constructing the prediction intervals is straightforward. First, a regression model is trained on a \emph{proper training} subset indexed by $\mathcal{I}_1 \subset \{1,\dots,n\}$ to predict $\hat{\mu}(x)$. Then, a disjoint \emph{calibration} subset indexed by $\mathcal{I}_2 \subset \{1,\dots,n\}$ is used to compute the non-conformity scores

\begin{equation}\label{eq:nonconf_score_scp}
    R_i = |Y_i-\hat{\mu}(X_i)| \quad i \in \mathcal{I}_2 \, ,
\end{equation}
i.e., the residuals. Computing
\begin{equation}\label{eq:empirical_quantile_scp}
    Q_{1-\alpha}(R, \mathcal{I}_2) := (1-\alpha)\left(1+ \frac{1}{|\mathcal{I}_2|}\right)\text{-th empirical quantile of } \left\{R_i\right\}_{i \in \mathcal{I}_2} \, ,
\end{equation}
with $|\mathcal{I}_2|$ the cardinality of the calibration set, yields the prediction interval
\begin{equation}\label{eq:prediction_intervals_scp}
    \mathcal{C}(X_{n+1}) = \left[\hat{\mu}(X_{n+1}) - Q_{1-\alpha}(R, \mathcal{I}_2),\, \hat{\mu}(X_{n+1}) + Q_{1-\alpha} (R, \mathcal{I}_2)\right] \, ,
\end{equation}
which satisfies \eqref{eq:coverage_guarantee}. Setting $\alpha = 0.1$ corresponds to 90\% prediction intervals.

The coverage property \eqref{eq:coverage_guarantee} is \textit{marginal}, i.e., averaged over all samples $\{(X_i, Y_i)\}_{i=1}^{n+1}$ \citep{romanoConformalizedQuantileRegression2019, feldmanImprovingConditionalCoverage2021}. The coverage guarantee holds for the distribution of the test point $X_{n+1}$, but may not hold for the conditional distribution of $Y_{n+1}$ given $X_{n+1}$,
\begin{equation}\label{conditional_coverate_property}
    \mathbb{P}\left\{Y_{n+1} \in \mathcal{C}(X_{n+1}) \,\vert\, X_{n+1} \right\} \geq 1-\alpha \, ,
\end{equation}
which is referred to as conditional coverage and in general, cannot be achieved \citep{lei2012distributionfreepredictionbands}. As a result, the coverage probability can fall below $100(1-\alpha)\%$ for subsets of the data, such as certain regimes or ranges of the target variable.

\subsection{Conformalized quantile regression (CQR)}\label{sec:conformalized_quantile_regression}
While SCP allows for constructing marginally valid prediction intervals in a straightforward way at almost no additional computational cost, Eq.~\eqref{eq:prediction_intervals_scp} reveals that the length of the prediction interval $\mathcal{C}(X_{n+1})$, is fixed to $2Q_{1-\alpha}(R, \mathcal{I}_2)$, independent of the input $X_{n+1}$. SCP is based on the implicit assumption that the spread of the residuals is constant for all inputs $X_{n+1}$, i.e. \textit{homoscedasticity}, which is in practice often not the case. To address this issue, \citet{romanoConformalizedQuantileRegression2019} proposed \textit{conformalized quantile regression} (CQR). In CQR, a quantile regression (QR) model is used to derive prediction intervals that can handle heteroscedasticity and thus are potentially more informative.

Quantile regression \citep{koenkerRegressionQuantiles1978} is a statistical method to estimate conditional quantiles of $Y_{n+1}$ given $X_{n+1}$, instead of the conditional mean. Here, a QR model is trained on the proper training dataset to predict the lower and upper quantile $\hat{Q}_{\alpha_\text{lo}}(x)$ and $\hat{Q}_{\alpha_\text{hi}}(x)$, which constitute initial estimates of the lower and upper bound of the prediction interval, $\hat{\mathcal{C}}(x) = [\hat{Q}_{\alpha_\text{lo}} (x), \hat{Q}_{\alpha_\text{hi}}(x)]$, with $\alpha_\text{lo} = \alpha/2$ and $\alpha_\text{hi} = 1-\alpha/2$, e.g. $\hat{\mathcal{C}}(x) = [\hat{Q}_{0.05}, \hat{Q}_{0.95}]$.
To quantify the error of this ad-hoc prediction interval, non-conformity scores
\begin{equation}\label{eq:nonconf_score_cqr}
    E_i := \max \{\hat{Q}_{\alpha_\text{lo}} (X_i) - Y_i,\, Y_i - \hat{Q}_{\alpha_\text{hi}} (X_i)\} \, ,
\end{equation}
are computed on the calibration dataset for each $i \in \mathcal{I}_2$. For a new input $X_{n+1}$, the prediction interval is conformalized by computing $Q_{1-\alpha}(E, \mathcal{I}_2)$ analogously to SCP \eqref{eq:empirical_quantile_scp}. This yields the conformalized prediction interval
\begin{equation}
    \mathcal{C}(X_{n+1}) = \left[\hat{Q}_{\alpha_\text{lo}} (X_{n+1}) - Q_{1-\alpha}(E, \mathcal{I}_2),\, \hat{Q}_{\alpha_\text{hi}}(X_{n+1}) + Q_{1-\alpha} (E, \mathcal{I}_2)\right] \, .
\end{equation}

\section{Data and methods}\label{sec:methods}
\subsection{Microphysical process rates}
We aim to reconstruct microphysical process rates from output of the ICON model, which were not included in the model output at the time of simulation, using a two-moment bulk microphysics scheme.

\begin{figure}
    \centering
    \includegraphics[width=\textwidth]{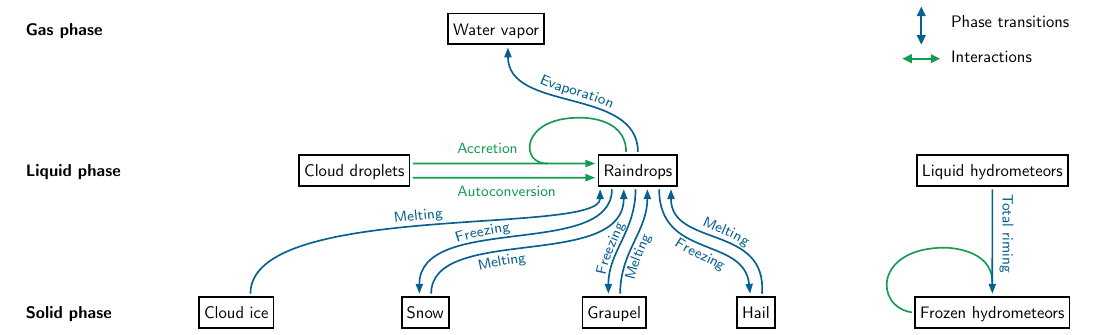}
    \caption{Schematic of selected microphysical processes (arrows) between the six hydrometeor categories (boxes) and water vapor in the two-moment microphysics scheme of \citet{seifertbehengTwomomentCloudMicrophysics2006}. Horizontal (green) arrows represent interaction processes, vertical (blue) arrows represent phase transitions. For simplicity, riming is shown separately. If a process occurs more than once, arrows represent contributions to the total process rate. From an ML perspective, boxes represent input features (mass mixing ratios and number concentrations) and arrows represent targets (process rates)\label{fig:microphysical_processes}}
\end{figure}

Bulk parameterization schemes describe cloud microphysical properties within each model grid volume with statistical bulk quantities, i.e. moments of the size distribution functions of distinct hydro\-meteor categories. The two-moment microphysics scheme \citep{seifertbehengTwomomentCloudMicrophysics2006} parameterizes cloud microphysics with prognostic number concentrations and mass mixing ratios of cloud droplets, raindrops, cloud ice, snow, graupel and hail, corresponding to the first two moments of the respective size distribution functions. MPRs describe a change in the prognostic variables\footnote{Here, we only consider process rates related to a change in the mass mixing ratio.} through phase transitions and interaction processes (Figure~\ref{fig:microphysical_processes}). Here, we consider two warm-rain processes, which both describe an increase in the rain mass mixing ratio through the formation of raindrops from coalescing cloud droplets (autoconversion) and cloud droplets and raindrops (accretion). Evaporation of rain causes a decrease in rain water mass. Rain mass also decreases by freezing of rain to snow, graupel and hail at temperatures $T < 0^\circ \,\text{C}$. At $T > 0^\circ \,\text{C}$, all frozen hydrometeors can melt to rain, leading to an increase in the rain mass mixing ratio. Furthermore, we consider riming, whereby liquid particles freeze upon contact with a frozen hydrometeor. The riming rate comprises eight individual processes between each category of hydrometeors in the liquid and solid phase. The melting, freezing and riming rates represent the sum of all individual processes. As inputs to the ML models, we use mass mixing ratios and number concentrations together with temperature, pressure and density\footnote{These are listed in the supplementary material~\ref{sec:icon_output_variables}. A short discussion of exchangeability can be found in the supplementary material~\ref{sec:exchangeability}.}.

\subsection{ICON model simulations and sampling procedure}
For training, calibration and evaluation, we use a dataset compiled from simulation output of the ICOsahedral Nonhydrostatic (ICON) model (version 2.6.6) \citep{zanglICONICOsahedralNonhydrostatic2015} in a limited-area configuration with $\overline{\Delta x} \approx 2$\,\text{km} effective grid spacing. The simulations are performed with the two-moment microphysics scheme \citep{seifertbehengTwomomentCloudMicrophysics2006} with a fast-physics timestep $\mathrm{t}_\mathrm{fast} = \mathrm{20\,s}$. The modeling domain extends from $-0.4^\circ\,$W to $17.7^\circ\,$E and from $43.7^\circ\,$N to $57.26^\circ\,$N, corresponding to the operational ICON-D2 configuration \citep{reinert2025DWDDatabase}. Each vertical column contains 65 levels up to 22 km above ground. The dataset comprises 19 one-day simulations (24 hours each) from January 2022 to July 2023 with an output time step of 10 minutes. We perform simulations for one day for each month. For training purposes, the MPRs are added as additional output variables in our setup. Further details on the model configuration are given in the supplementary material~\ref{sec:icon_configuration}. The set of input features for each target variable is aligned with the computation of the process rate in the ICON subroutine and listed in the supplementary material~\ref{sec:input_features}. The size of the initial dataset is reduced by random sampling of model grid points. Furthermore, the output from high altitudes is discarded under the assumption that they are cloud-free. We use the simulation output for 2022 for training (even-numbered months) and validation (odd-numbered months), the simulated days for 2023 are used for the test dataset. The training dataset is randomly split into the proper training and the calibration set. The datasets for each target MPR are retrieved from these initial datasets by filtering out all grid points where the value of the MPR fulfills $\left|\mathrm{MPR}\right| \geq 10^{-12} \, \text{kg}\,\text{kg}^{-1}\,{\text{t}_\text{fast}}^{-1}$ and $\sum_k q_k  \geq 10^{-12}\, \text{kg}\,\text{kg}^{-1}$ over all mass mixing ratios $q_k$ in the set of input features, which corresponds to the threshold $q_\text{crit} = 10^{-12}\, \text{kg}\,\text{kg}^{-1}$ in the two-moment microphysics scheme\footnote{In the ICON model, $q_\text{crit}$ is a critical mass mixing ratio threshold used in the parameterization of certain processes, such as autoconversion.}. This allows us to train the regression models to predict logarithmically transformed targets, as the range of the target variables spans multiple orders of magnitude. Thus, the proper training dataset contains $1.12 \times 10^7$ samples and each validation, calibration and evaluation dataset contains $1.6 \times 10^6$ samples, corresponding to a 70\%/10\%/10\%/10\% split. The input features are scaled to the range $[0, 1]$ with \textit{min-max} scaling.

\subsection{Training and calibration}
A separate model is trained for each process rate. In order to obtain prediction intervals with SCP, we train random forest (RF), gradient boosting (XGB) and neural network (NN) models to obtain point-predictions for the process rates. For this, we use the \texttt{scikit-learn} library for the RF models \citep{pedregosaScikitlearnMachineLearning2011}, the \texttt{xgboost}  library \citep{chenXGBoostScalableTree2016} for the XGB models and \texttt{pytorch} \citep{paszkePyTorchImperativeStyle2019} for the NNs. Calibration is performed with the absolute error residual \eqref{eq:nonconf_score_scp} and $\alpha = 0.1$. For CQR, we employ three QR architectures: quantile RF (QRF) models using \texttt{quantile\_forest}'s \textit{RandomForestQuantileRegressor} \citep{meinshausenQuantileRegressionForests2006, johnsonQuantileForest2024}, quantile gradient boosting (QXGB) and quantile feed-forward neural networks (QNN). We use the Adam optimizer \citep{kingmaAdamMethodStochastic2017} for training the QNN with a learning rate scheduler with a minimum learning rate of $10^{-6}$. The maximum number of epochs is 150. The QNN is trained with the \textit{quantile loss} \citep{koenkerRegressionQuantiles1978}, frequently also referred to as the \textit{pinball loss}, 
\begin{equation}
    L_\alpha (y, \hat{y}) = \alpha (y - \hat{y}) \vmathbb{1} \left\{y > \hat{y}\right\} + (1 - \alpha)(\hat{y} - y)\vmathbb{1} \left\{y \leq \hat{y}\right\} \, , 
    \label{eq:pinball_loss}
\end{equation}
where $ \hat{y}$ is the model's prediction of the ``true'' $y$. All three QR models are trained to predict the $5\%$- and $95\%$-quantiles. More details on the training and the model hyperparameters are given in the supplementary material~\ref{sec:hyperparameters}. The predicted upper and lower quantiles are calibrated with the non-conformity score function \eqref{eq:nonconf_score_cqr} and $\alpha = 0.1$.

\section{Results}\label{sec:results}
\subsection{Deterministic performance of microphysical process rate emulation}
The performance of the deterministic RF, XGB and NN models is evaluated with the coefficient of determination $R^2$ (Table~\ref{table:results_point_predictions}). In order to obtain an estimate of the performance of the QR models, we compute the $R^2$ score with the median of the predicted (uncalibrated) upper and lower quantile. 
\begin{table*}[t]
    \tabcolsep=0pt%
    \centering
    \caption{Deterministic performance in terms of the $R^2$ score. The scores for the QR models are computed with the median of the predicted uncalibrated upper and lower quantile. For each process rate, the highest score for the deterministic and QR models is highlighted in bold\label{table:results_point_predictions}}
    \begin{tabular*}{\textwidth}{@{\extracolsep{\fill}}lccccccc@{}}
    \toprule
    {}   & Autoconversion & Accretion     & Rain evap.    & Rain fr.      & Rain melt.    & Tot. riming   \\
    \midrule
    RF   & 0.67           & 0.97          & 0.93          & 0.52          & 0.90          & 0.78          \\
    XGB  & 0.66           & 0.70          & 0.81          & 0.47          & 0.97          & 0.90          \\
    NN   & \textbf{0.77}  & \textbf{0.98} & \textbf{0.98} & \textbf{0.78} & \textbf{0.99} & \textbf{0.91} \\
    \midrule
    QRF  & 0.02   & \textbf{0.99}    & \textbf{0.98}     & 0.45   & 0.98     & \textbf{0.91}       \\
    QXGB & \textbf{0.60} & 0.74    & 0.83     & 0.46   & 0.88     & 0.82       \\
    QNN  & 0.43   & 0.88    & 0.66     & \textbf{0.68}   & \textbf{0.99}     & 0.82      \\
    \botrule
    \end{tabular*}
\end{table*}
The deterministic models show moderate to high $R^2$ scores ranging from $0.47$ to $0.99$, with the NN models achieving the highest $R^2$ scores for all MPRs. The lowest score is obtained for rain freezing with the XGB model. The QR models yield $R^2$ scores that are largely comparable to those of the deterministic models. An exception is the QRF model for autoconversion, where $R^2 \approxeq 0.02$. For all other QR models and process rates, $R^2$ ranges from $0.43$ to $0.99$.

\subsection{Conformal prediction uncertainty estimates}
We evaluate the calibration of the CP intervals with the prediction interval coverage probability (PICP)
\begin{equation}
    \text{PICP} = \frac{1}{n} \sum_{i=1}^n \vmathbb{1} \left\{Y_i \in \mathcal{C}(X_{i})\right\} \, ,
    \label{eq:picp}
\end{equation}
which is the proportion of the true values that are covered by the prediction intervals, i.e. the PICP evaluates the marginal coverage property \eqref{eq:coverage_guarantee}. Here, $n$ is the size of the test set, $Y_i$ the true value and $\mathcal{C}(X_{i})$ the prediction interval. With $\alpha = 0.1$, we aim for $\text{PICP} \approx 90\%$. The results are given in Table~\ref{table:results_picp}; for comparison, we include the PICP for the uncalibrated prediction intervals obtained with the QR models. 
\begin{table*}[t]
    \tabcolsep=0pt%
    \centering
    \caption{Prediction interval coverage probability (PICP). For comparison, we show the results for the uncalibrated prediction intervals obtained with QR in gray. For each process rate, the best PICP obtained with SCP and CQR is highlighted in bold, which is defined as $\min (\lvert \mathrm{PICP} - 90\%\rvert)$\label{table:results_picp}} 
    \begin{tabular*}{\textwidth}{@{\extracolsep{\fill}}lccccccc@{}}
    \toprule
    {}        & Autoconversion  & Accretion        & Rain evap.       & Rain fr.         & Rain melt.       & Tot. riming      \\
    \midrule
    SCP, RF       & 90.18\%          & 91.96\%          & \textbf{89.21\%} & 88.01\%          & 90.96\%          & 91.21\%          \\
    SCP, XGB      & 89.82\%          & \textbf{90.63\%} & 87.61\%          & 89.52\%          & 90.17\%          & 91.06\%          \\
    SCP, NN       & \textbf{89.98\%} & 91.43\%          & 87.61\%          & \textbf{90.47\%} & \textbf{89.98\%} & \textbf{90.74\%} \\
    \midrule
    \textcolor{gray}{QRF, uncalibrated} & \textcolor{gray}{89.47\%} & \textcolor{gray}{88.46\%} & \textcolor{gray}{97.91\%} & \textcolor{gray}{87.61\%} & \textcolor{gray}{94.53\%} & \textcolor{gray}{97.71\%} \\
    CQR, QRF  & 86.44\% & 88.51\% & {\textbf{90.05\%}}  & 87.61\% & {89.18\%}  & {88.63\%}   \\
    \textcolor{gray}{QXGB, uncalibrated} & \textcolor{gray}{89.24\%} & \textcolor{gray}{89.50\%} & \textcolor{gray}{89.58\%} & \textcolor{darkgray}{\textbf{90.66\%}} & \textcolor{gray}{89.00\%} & \textcolor{darkgray}{\textbf{90.28\%}} \\
    CQR, QXGB & 89.27\% & 88.83\% & 88.14\%  & 89.18\% & 88.09\%  & 89.14\%    \\
    \textcolor{gray}{QNN, uncalibrated} & \textcolor{gray}{89.68\%} & \textcolor{gray}{87.19\%} & \textcolor{gray}{95.71\%} & \textcolor{gray}{83.74\%} & \textcolor{gray}{90.87\%} & \textcolor{gray}{92.60\%} \\
    CQR, QNN  & {\textbf{89.77\%}} & \textbf{90.41\%} & 90.11\%  & {89.04\%} & \textbf{90.60\%}  & 89.43\%   \\
    \botrule
    \end{tabular*}
\end{table*}
We find that the $\text{PICP} \approx 90\%$ for both the deterministic models calibrated with SCP and the CQR models. Yet, in some cases, the PICP falls slightly below the target $90\%$. This may be due to sampling variability but also could indicate mild violations of the exchangeability assumption. Furthermore, we observe that calibration does not always result in a PICP that is closer to 90\%, compared to the uncalibrated intervals. This can be seen, for instance, for autoconversion and the QRF model or accretion and the QXGB model. In these cases, calibration decreases the PICP slightly. This can occur because the CQR non-conformity score function \eqref{eq:nonconf_score_cqr} can be negative and thus shrink the predicted interval, as the score function is designed to account for both undercoverage and overcoverage \citep{romanoConformalizedQuantileRegression2019}. Nevertheless, in other cases we observe a significant improvement, for example with the QRF models for the melting to rain and total riming rate.

The sharpness of the prediction intervals is evaluated with the normalized mean prediction interval width (NMPIW)
\begin{equation}\label{eq:NMPIW}
    \mathrm{NMPIW} = \frac{1}{n} \sum_{i=1}^n \left(\frac{U(X_i) - L(X_i)}{\max(Y) - \min(Y)}\right) \, ,
\end{equation}
where $U$ and $L$ refer to the calibrated upper and lower prediction interval bounds. We present the results in Table~\ref{table:results_al}\footnote{The length of the calibrated and uncalibrated prediction intervals is similar in each case, therefore, we do not include both results.}.
\begin{table*}[t]
    \tabcolsep=0pt%
    \centering
    \caption{Normalized mean prediction interval width (NMPIW). For each process rate, the smallest NMPIW obtained with SCP and CQR, is highlighted in bold}
    \label{table:results_al}
    \begin{tabular*}{\textwidth}{@{\extracolsep{\fill}}lcccccc@{}}
    \toprule
    {} & Autoconversion              & Accretion          & Rain evap.             & Rain fr.              & Rain melt. & Tot. riming \\
    \midrule
    SCP, RF & $4.89\times 10^{-6}$        & $1.23\times 10^{-4}$ & $8.64\times 10^{-4}$ & $\mathbf{1.66\times 10^{-8}}$ & $2.79\times 10^{-3}$  & $ 4.22\times 10^{-4}$ \\
    SCP, XGB & $4.29\times 10^{-6}$       & $\mathbf{4.88\times 10^{-5}}$ & $3.21\times 10^{-4}$ & $ 2.97\times 10^{-8}$ & $\mathbf{9.04\times 10^{-4}}$  & $\mathbf{1.38\times 10^{-4}}$ \\
    SCP, NN & $\mathbf{4.06\times 10^{-6}}$        & $7.31\times 10^{-5}$ & $\mathbf{1.80\times 10^{-4}}$ & $ 2.96\times 10^{-8}$ & $1.08\times 10^{-3}$  & $ 1.94\times 10^{-4}$ \\
    \midrule
    CQR, QRF         & $\mathbf{8.86\times 10^{-5}}$ & $\mathbf{1.08\times 10^{-4}}$ & $7.49\times 10^{-4}$ & $5.62\times 10^{-5}$ & $\mathbf{1.48\times 10^{-3}}$  & $\mathbf{4.62\times 10^{-4}}$ \\
    CQR, QXGB        & $1.12\times 10^{-4}$ & $3.17\times 10^{-4}$ & $\mathbf{7.36\times 10^{-4}}$ & $7.46\times 10^{-5}$ & $2.30\times 10^{-3}$  & $6.28\times 10^{-4}$ \\
    CQR, QNN         & $1.10\times 10^{-4}$ & $1.21\times 10^{-4}$ & $9.13\times 10^{-4}$ & $\mathbf{3.60\times 10^{-5}}$ & $1.50\times 10^{-3}$  & $7.16\times 10^{-4}$ \\
    \botrule
    \end{tabular*}
\end{table*}
We find that for each process rate, the smallest NMPIW is obtained with SCP. For accretion, rain evaporation, rain melting and total riming, the NMPIW values obtained with SCP with the deterministic models and CQR with the QR models differ by not more than one order of magnitude for each MPR. Yet, for autoconversion and rain freezing, the NMPIW obtained with SCP is smaller by up to two and three orders of magnitude, respectively. Moreover, a higher PICP does not necessarily imply a larger or smaller NMPIW, as can be seen, e.g., for the rain freezing and the total riming rate, where the smallest NMPIW values correspond to neither the highest nor lowest PICP.

\begin{figure}[t]
    \centering
    \FIG{
    \includegraphics[align=t, width=0.48\textwidth]{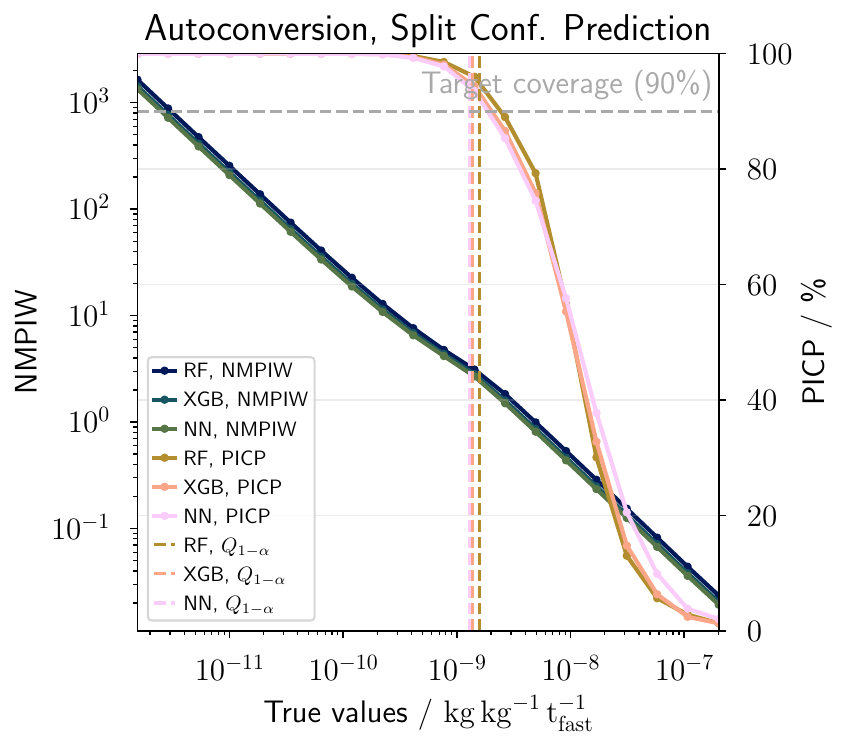}\hspace{0.02\textwidth}
    \includegraphics[align=t, width=0.48\textwidth]{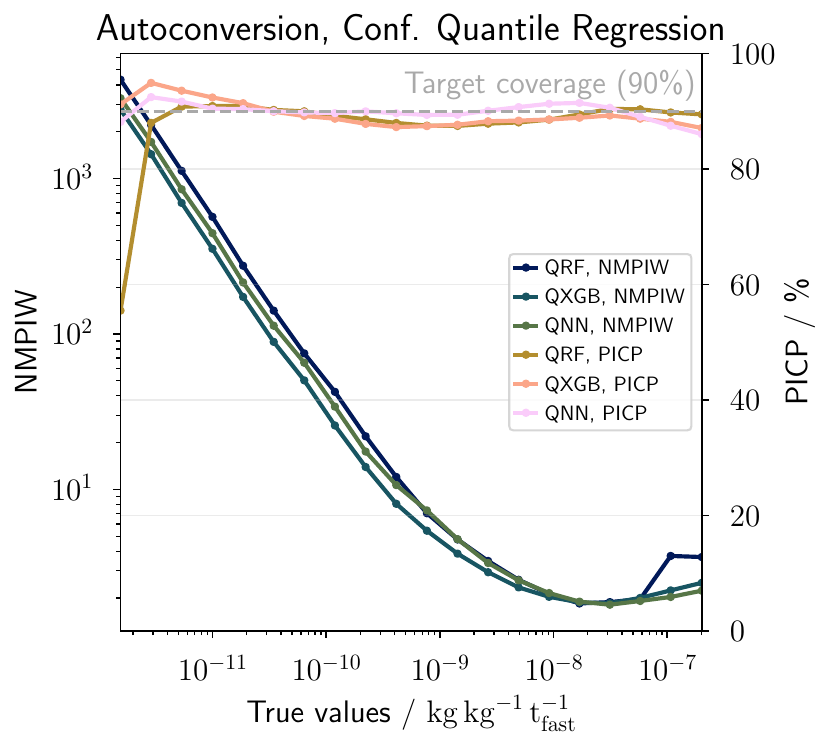}
    }
    \caption{Normalized mean prediction interval width (NMPIW) and prediction interval coverage probability (PICP) binned by value of the autoconversion rate for SCP (left) and CQR (right)} 
    \label{fig:plots_coverage}
\end{figure}

We visualize the prediction intervals derived with SCP and CQR exemplarily for autoconversion in Figure~\ref{fig:plots_intervals}\footnote{Additional results for all process rates are included in the supplementary material~\ref{sec:additional_results}.}.
\begin{figure}[t]
    \centering
    \FIG{\includegraphics[align=t, width=0.45\textwidth]{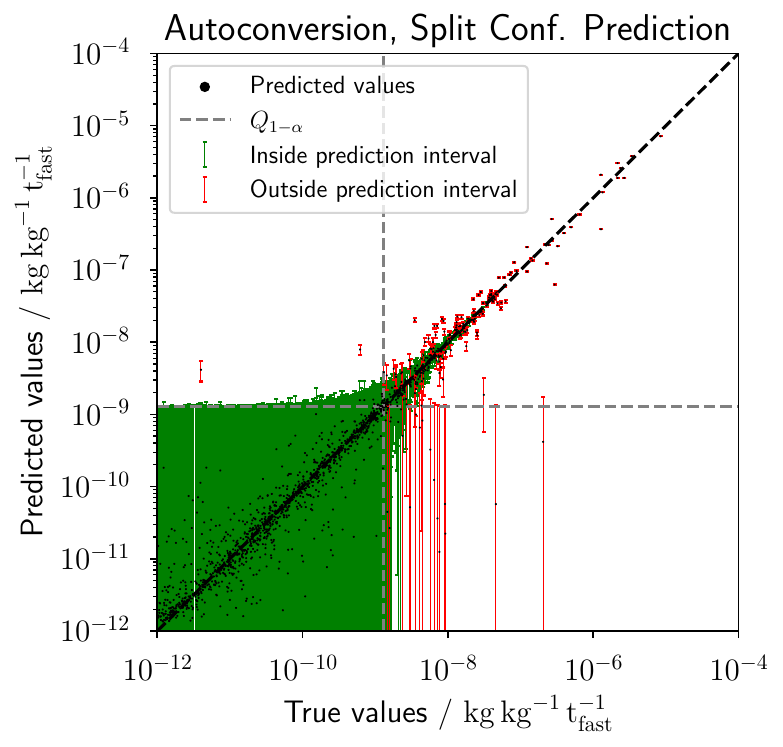}\hspace{0.05\textwidth}
    \includegraphics[align=t, width=0.45\textwidth]{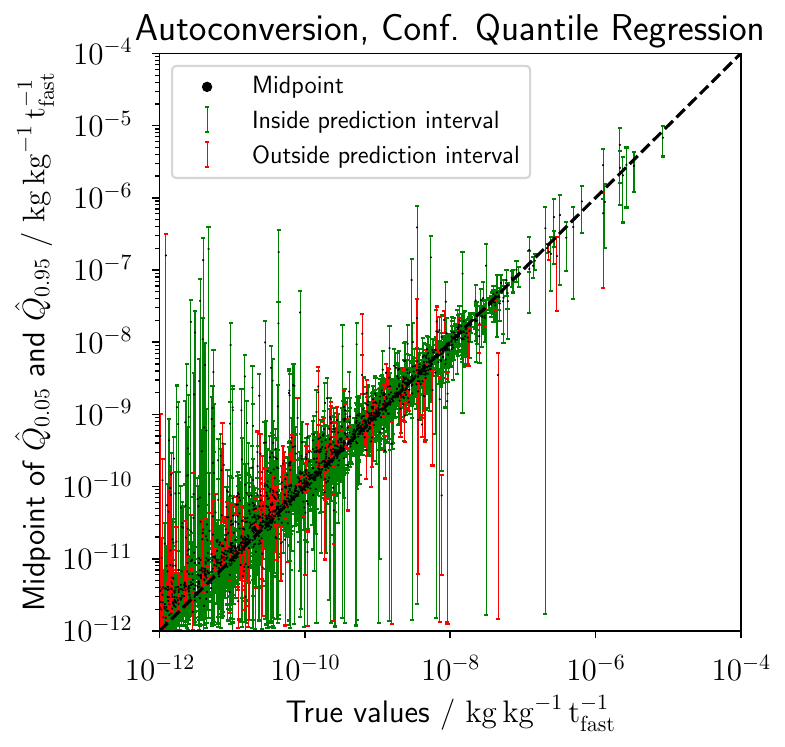}}
    {\caption{Calibrated prediction intervals with SCP and the NN (left) and CQR and the QNN (right) for autoconversion. For better visualization, we only show 1500 randomly selected samples}
    }
    \label{fig:plots_intervals}
\end{figure}
It is apparent that the SCP results mainly depend on $Q_{1-\alpha}$. For true values smaller than $Q_{1-\alpha}$, all prediction intervals have the same width and mostly cover the true value. Above $Q_{1-\alpha}$, this is not the case anymore, with prediction intervals with decreasing PICP typically missing the true value of the MPR. In contrast, with CQR, valid prediction intervals are present across the full range of true values. This indicates that in our case, single-valued evaluation metrics might not be reflective of the true behavior. To further investigate the dependence of the PICP and NMPIW on the size range of the target variables, in Figure~\ref{fig:plots_coverage} we show both quantities binned by the size of the true value for SCP and CQR, respectively.
Consistent with Figure~\ref{fig:plots_intervals}, we observe that with SCP, the PICP is almost at 100\% for size bins of the true values below $Q_{1-\alpha}$ and drops to $\text{PICP} \ll 90\%$ above $Q_{1-\alpha}$. With CQR, the PICP is approximately 90\% across all size bins, in agreement with Figure~\ref{fig:plots_intervals}. With both methods, the NMPIW decreases with increasing size of the target value. Thus, unlike SCP, CQR can predict state-dependent intervals across the entire range of MPRs, yielding more reliable prediction intervals in general\footnote{The supplementary material~\ref{sec:heteroscedasticity} includes an analysis of the absolute residuals, which is a diagnostic for heteroscedasticity.}.

\section{Conclusions}\label{sec:conclusions}
We applied and evaluated conformal methods to derive calibrated prediction intervals for microphysical process rates (MPRs), enabling the retrieval of detailed process information from high-resolution atmospheric simulations. We compared split conformal prediction, which is simple and inexpensive, with conformalized quantile regression, which requires training a quantile regression model. Applied to well-performing machine learning emulators, both approaches yield satisfactory results for calibration (measured with PICP) and sharpness (measured with NMPIW) on average. In certain cases, the PICP falls slightly below the nominal level, which might indicate mild violations of the exchangeability assumption. Moreover, split conformal prediction exhibits a strongly bin-dependent PICP and unreliable intervals for process rate values exceeding $Q_{1-\alpha}$. In contrast, conformalized quantile regression adapts to heteroscedasticity and thus better reflects the distribution of MPRs, making it the preferable approach when large values are most important. Future work could explore alternative non-conformity scores \citep{ leiDistributionFreePredictiveInference2018, gopakumarUncertaintyQuantificationSurrogate2025, papadopoulosNormalizedNonconfomityMeasures2008} and loss functions for quantile regression training \citep[e.g.,][]{feldmanImprovingConditionalCoverage2021} to further improve efficient, reliable uncertainty quantification across the diverse distributions encountered in weather and climate science.

\begin{Backmatter}

\paragraph{Acknowledgments}
The authors gratefully acknowledge the computing time provided on the high-performance computer HoreKa by the National High-Performance Computing Center at KIT (NHR@KIT). This center is jointly supported by the Federal Ministry of Education and Research and the Ministry of Science, Research and the Arts of Baden-Württemberg, as part of the National High-Performance Computing (NHR) joint funding program (\url{https://www.nhr-verein.de/en/our-partners}). HoreKa is partly funded by the German Research Foundation (DFG).
The GitHub Copilot extension for Visual Studio Code and ChatGPT-5.2 was used to support the development of the model code. Generative AI tools have only been used for proofreading; not for text preparation. Figure~\ref{fig:plots_coverage}, Figure~\ref{fig:scores_scp_app}, Figure~\ref{fig:scores_cqr_app}, Figure~\ref{fig:residuals_scp} and Figure~\ref{fig:residuals_cqr} make use of the Scientific color maps 7.0 \citep{crameriMisuseColourScience2020}. Figure~\ref{fig:cp_algorithm} and Figure~\ref{fig:microphysical_processes} were created using the \LaTeX package Ti\textit{k}Z \citep{tikzTantau2013}. 

\paragraph{Author Contributions}
Conceptualization: M.S., T.B., C.H., Data curation: M.S:, Data visualization: M.S., Methodology: M.S.,  Writing original draft: M.S., Editing: M.S., T.B., C.H.. All authors approved the final submitted draft.

\paragraph{Competing Interests}
The authors declare none. 

\paragraph{Data Availability Statement}
The released version of the data and code necessary to reproduce the manuscript's figures are publicly available on Zenodo  at \url{https://zenodo.org/records/19114006} (DOI: \url{https://doi.org/10.5281/zenodo.19114006}) and in the GitHub repository \url{https://github.com/miriamsimm/ConformalMPR}. The trained models as well as the ICON model simulation output are available upon request. The newest version of the source code that was used to perform the simulations with the ICON model can be found at \url{icon-model.org}.

\paragraph{Ethical Standards}
The research meets all ethical guidelines, including adherence to the legal requirements of the study country.

\paragraph{Funding Statement}
M.~S. and C.~H. acknowledge funding from the NHR Call for Collaboration Project MICRO. C.~H. acknowledges funding from the European Union's Horizon Europe Programme under Grant Agreement No. 101137639 (CleanCloud). T.~B. acknowledges funding from the Swiss State Secretariat for Education, Research and Innovation (SERI) for the Horizon Europe project AI4PEX (Grant agreement ID: 101137682 and SERI no 23.00546).

\nocite{simm_2026_19114006}
\printbibliography

\begin{appendix}
\section*{Supplementary Material}
\section{Simulations with the ICOsahedral Nonhydrostatic (ICON) model}
\subsection{ICON model configuration}\label{sec:icon_configuration}
In Table~\ref{tab:icon_model_configuration}, we give details about the model configuration used for simulations with the ICON model \citep{zanglICONICOsahedralNonhydrostatic2015}, for which we use the NWP physics package \citep{prillWorkingWithTheICONModel2024}.
\begin{table*}[ht]
    \centering
    \caption{Configuration of the ICON model simulations with the NWP physics package \citep{prillWorkingWithTheICONModel2024}\label{tab:icon_model_configuration}}
    \begin{tabular*}{\textwidth}{@{\extracolsep{\fill}}ll@{}}
        \toprule
            {Model aspect} & {Setting} \\
        \midrule
            Model grid & {R19B07} \\
            Vertical coordinates & {Height based terrain-following smooth level vertical} \\
            {} & {(SLEVE) coordinate system \citep{leuenbergerAGeneralizationSLEVEVerticalCoordinate2010}} \\
            Initial and boundary data & ICON-EU analyses, {3\,h} update \\
            Initialization time        & 00:00 UTC \\
            Integration time          & {24\,hours} \\
            Turbulence scheme & Prognostic TKE \citep{raschendorferNewTurbulenceParameterization2001} \\
            Microphysics scheme & Two-moment microphysics \citep{seifertbehengTwomomentCloudMicrophysics2006} \\
            Convection scheme & Explicit deep convection, parameterized shallow convection \\
            {} &  \citep{bechtoldAdvancesSimulatingAtmospheric2008, tiedtkeAComprehensiveMassFluxScheme1989} \\
            Cloud condensation nuclei activation &  Segal-Khain scheme \citep{segalDependenceDropletConcentration2006} \\
            Land-surface model & Multilayer land-surface scheme TERRA \\ 
            {} & \citep{schrodinheiseLandSurfaceModel2001} \\
            Radiation scheme & ecRad \citep{bozzohoganAFlexibleEfficientRadiationScheme2018}\\
        \botrule
    \end{tabular*}
\end{table*}

\subsection{ICON output variables}\label{sec:icon_output_variables}
In Table~\ref{tab:icon_output_variables}, we provide a list of the ICON output variables together with their respective unit and variable name.
\begin{table*}[ht]
    \centering
    \caption{ICON output variables\label{tab:icon_output_variables}}
    \begin{tabular}{c l l}
        \toprule
            {Variable} & {Unit} & {Description} \\
        \midrule
            {$q_c$} & {kg kg$^{-1}$} & {Specific cloud water content} \\
            {$q_r$} & {kg kg$^{-1}$} & {Specific rain content} \\
            {$q_i$} & {kg kg$^{-1}$} & {Specific cloud ice content} \\
            {$q_s$} & {kg kg$^{-1}$} & {Specific snow content} \\
            {$q_g$} & {kg kg$^{-1}$} & {Specific graupel content} \\
            {$q_h$} & {kg kg$^{-1}$} & {Specific hail content} \\
            {$n_c$} & {kg$^{-1}$} & {Cloud droplet number concentration} \\
            {$n_r$} & {kg$^{-1}$} & {Rain drop number concentration} \\
            {$n_i$} & {kg$^{-1}$} & {Cloud ice number concentration} \\
            {$n_s$} & {kg$^{-1}$} & {Snow number concentration} \\
            {$n_g$} & {kg$^{-1}$} & {Graupel number concentration} \\
            {$n_h$} & {kg$^{-1}$} & {Hail number concentration} \\
            {$q_v$} & {kg kg$^{-1}$} & {Specific humidity} \\
            {$\rho$} & {kg m$^{-3}$} & {Density} \\
            {$p$} & {Pa} & {Pressure} \\
            {$T$} & {K} & {Temperature} \\
        \botrule
    \end{tabular}
\end{table*}

\subsection{Exchangeability}\label{sec:exchangeability}
A core assumption of CP is that the data is exchangeable. In weather and climate modeling, this is not trivial, as the data often has spatio-temporal structure. Exchangeability is violated if nearby data points are correlated \citep{angelopoulosGentleIntroductionConformal2022,barber2023conformalpredictionexchangeability}.
In this work, we seek to obtain 90\% prediction intervals that contain the value of the MPR as if the process rate would have been included in the ICON model output. Thus, we are concerned with the mapping of the standard ICON output variables (mass mixing ratios $q_k$, number concentrations $n_k$, etc.) to the values of the MPRs. This mapping is time-independent. In ICON, the process rates are a function of the mass mixing ratios and number concentrations of the hydrometeor classes as well as atmospheric state variables. Although these variables itself have non-local dependencies, due to, for instance, temperature gradients and updraft, the internal computation of the process rates in ICON does only depend on prognostic variables in the same model grid cell and does not depend on time. Thus, we assume that our data is exchangeable.

\section{Machine learning setup}
\subsection{Input features and targets}\label{sec:input_features}
In Table~\ref{tab:input_output}, we list the set of input features used for the emulation of each MPR. The selection of features is based on the prognostic cloud variables that are used in the each ICON subroutine to compute the respective process rate.
\begin{table}[ht]
    \centering
    \caption{Microphysical process rates (targets) and input features\label{tab:input_output}}
    \begin{tabular}{ll} 
        \toprule
        {Microphysical process rate} & {Input features} \\
        \midrule
        Autoconversion & {$q_c,\,q_r,\,n_c,\,n_r,\,q_v,\,\rho$}  \\
       Accretion & {$q_c,\,q_r,\,n_c,\,n_r,\,q_v,\,\rho$} \\
       Rain evaporation & {$q_c,\,q_r,\,n_r,\,q_v,\,\rho,\,T,\,p$} \\
        Melting to rain & {$q_c,\,q_r,\,q_s,\,q_g,\,q_h,\,n_c,\,n_r,\,n_s,\,n_g,\,n_h,\,q_v,\,\rho,\,T$} \\
        Rain freezing & {$q_r,\,n_r,\,T$} \\
         Total riming & {all $q_k$ and $n_k$, $q_v,\,\rho,\,T$} \\ 
        \botrule
    \end{tabular}
\end{table}%

\subsection{ML model hyperparameters}\label{sec:hyperparameters}
We list the hyperparameters of the deterministic random forest model (RF), gradient boosting model (XGB) and neural network (NN) and the quantile random forest (QRF), quantile gradient boosting model (QXGB) and quantile neural network (QNN) in Table~\ref{tab:hyperparameters}.
\begin{table}[ht]
    \centering
    \caption{ML model hyperparameters\label{tab:hyperparameters}} 
    \begin{tabular*}{\textwidth}{@{\extracolsep{\fill}}ll ll ll@{}}
        \toprule
        \multicolumn{2}{c}{RF \& QRF} 
        & \multicolumn{2}{c}{XGB \& QXGB} 
        & \multicolumn{2}{c}{NN \& QNN} 
        \\ 
        \midrule
        {Num. estimators}     & {150}  & {Num. estimators}         & {500}  & {Num. nodes} & {192}\\
        {Max. depth}         & {70}   & {Max. depth}             & {10}   & {Num. layers} & {4}\\
        {Min. samples split} & {2.0}  & {Learning rate}          & {0.02} & {Learning rate} & {0.001}\\
        {Min. samples leaf}  & {2.0}  & {Gamma}                  & {0.2}  & {Weight decay} & {$10^{-6}$}\\
        {Max. features}      & {1.0}  & {Min. child weight}      & {2.0}  & {Batch size} & {256}\\
        {Max. samples}       & {0.75} & {Subsample ratio}        & {1.0}  & {Activation function} & {ReLU}\\
        {}                   & {}     & {Column subsample ratio} & {1.0}  & {Max. epochs} & {150}\\
        {}                   & {}     & {L1 regularization}      & {0.1}  & {Optimizer} & {Adam}\\
        {}                   & {}     & {L2 regularization}      & {0.1}  & {NN Loss} & {MSE}\\
        {}                   & {}     & {}                       & {}     & {QNN Loss} & {Quantile loss}\\
        \botrule
    \end{tabular*}
\end{table}%

\section{Additional results}\label{sec:additional_results}
\subsection{Prediction intervals}\label{sec:pred_intervals_app}
In Figure~\ref{fig:intervals_scp_app} and Figure~\ref{fig:intervals_cqr_app}, we show the prediction intervals derived with SCP and CQR, respectively, for the accretion, rain evaporation, rain freezing, melting to rain and the total riming rate.
\begin{figure}[ht]
    \centering
    \FIG{\includegraphics[align=t, width=0.33\textwidth]{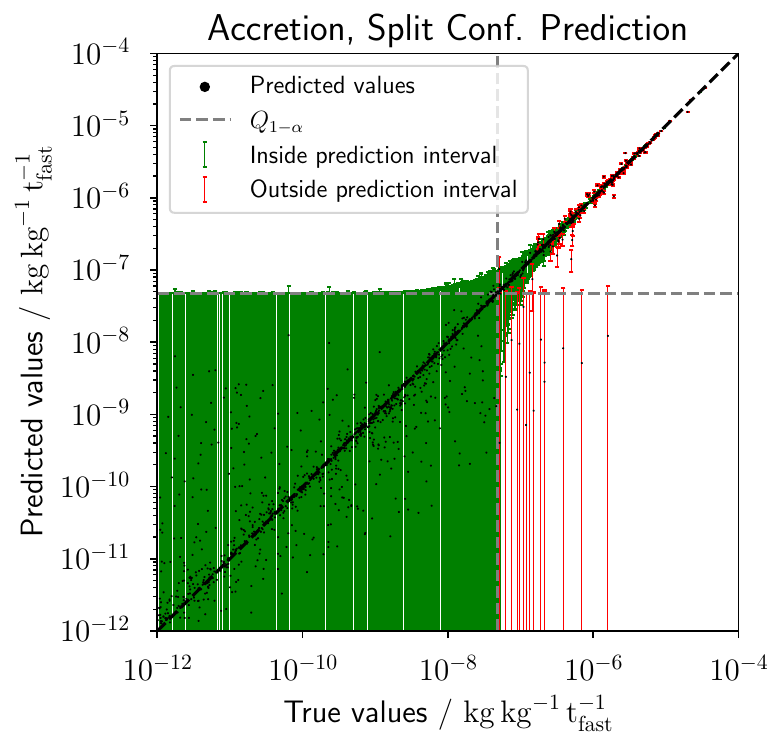}
    \includegraphics[align=t, width=0.33\textwidth]{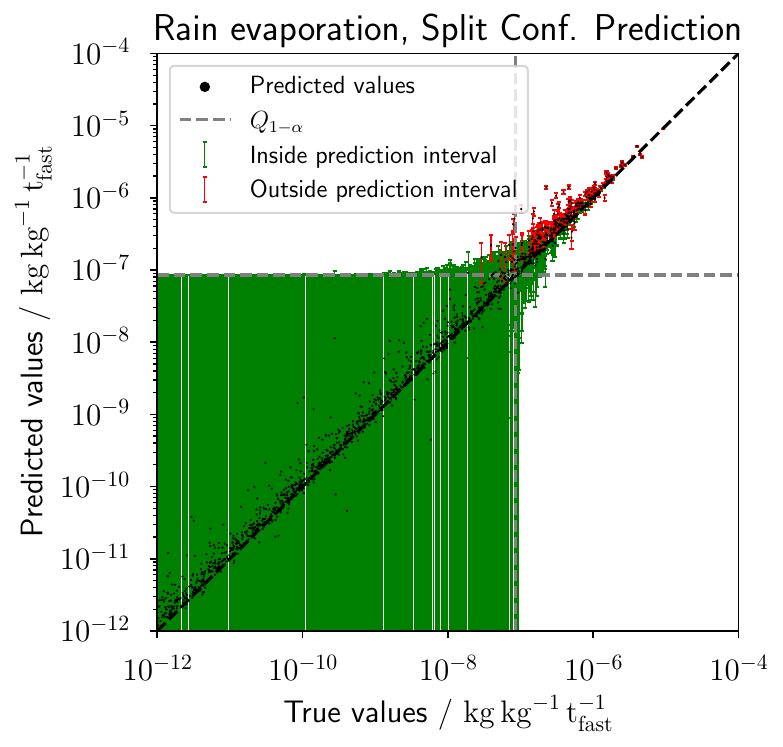} 
    \includegraphics[align=t, width=0.33\textwidth]{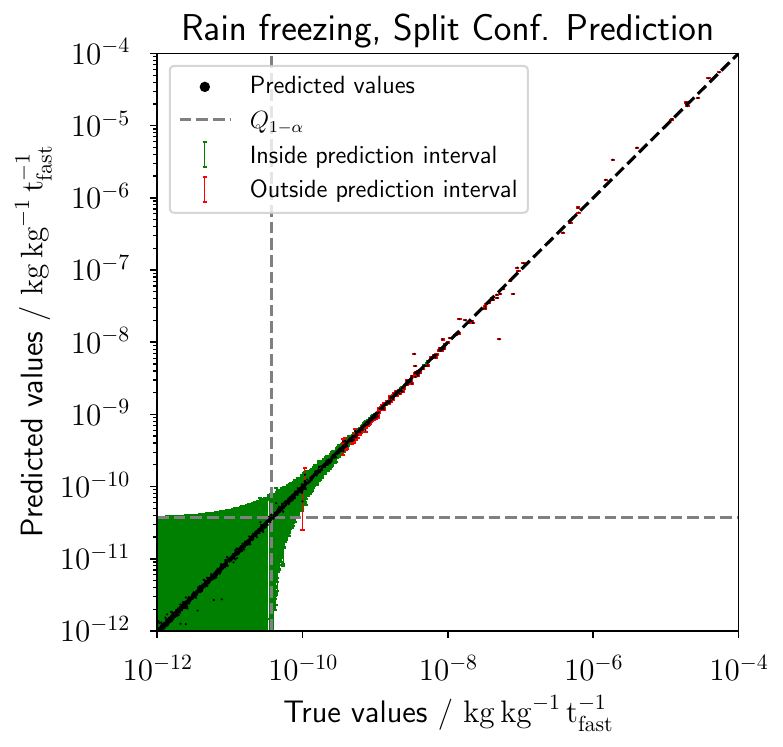}}
    \FIG{
    \includegraphics[width=0.33\textwidth]{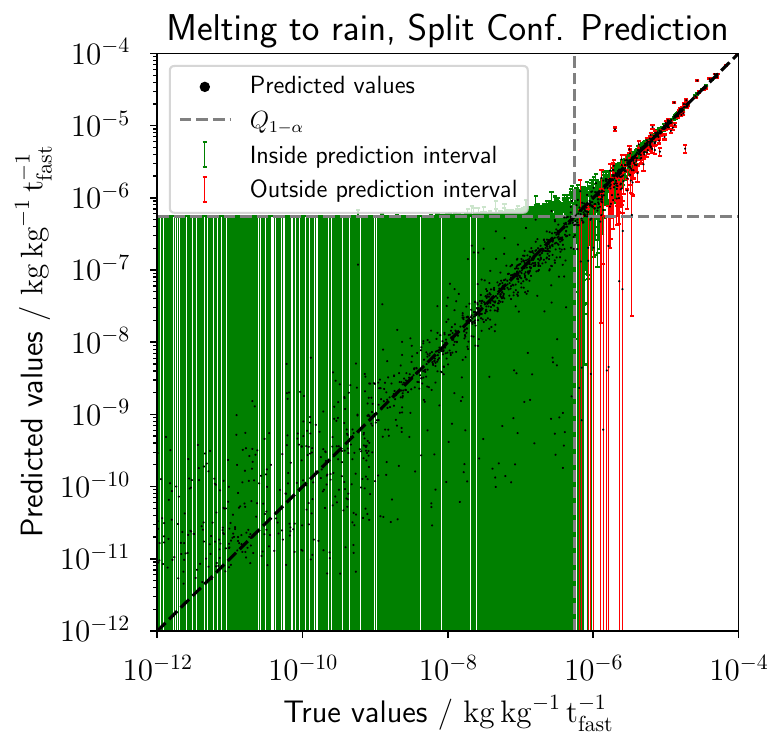}
    \includegraphics[width=0.33\textwidth]{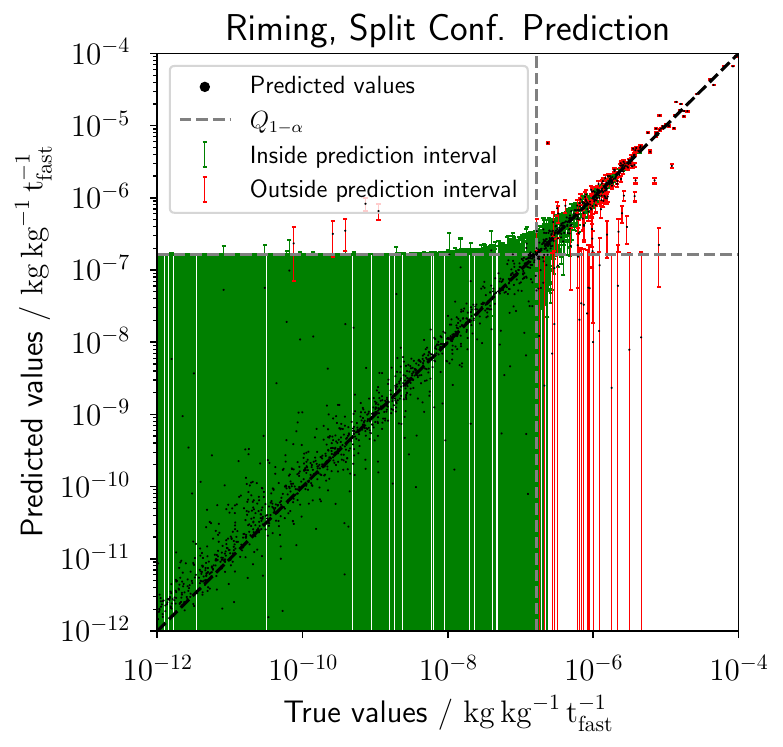} 
    }
    {\caption{Calibrated prediction intervals with split conformal prediction (SCP) for the accretion (NN), rain evaporation (RF), rain freezing (NN), melting to rain (NN) and total riming (NN) rate, obtained with the deterministic model that yields the best PICP (in brackets, see Table~\ref{table:results_picp} in the main text). For better visualization, we only show 1500 randomly selected samples}\label{fig:intervals_scp_app}}
\end{figure}
\begin{figure}
    \centering
    \FIG{\includegraphics[align=t, width=0.33\textwidth]{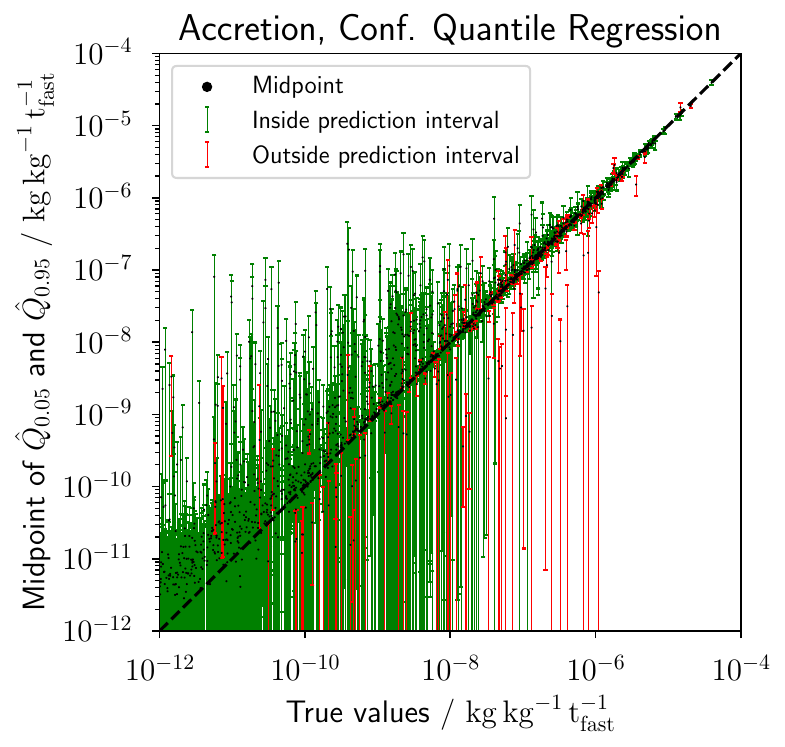}
    \includegraphics[align=t, width=0.33\textwidth]{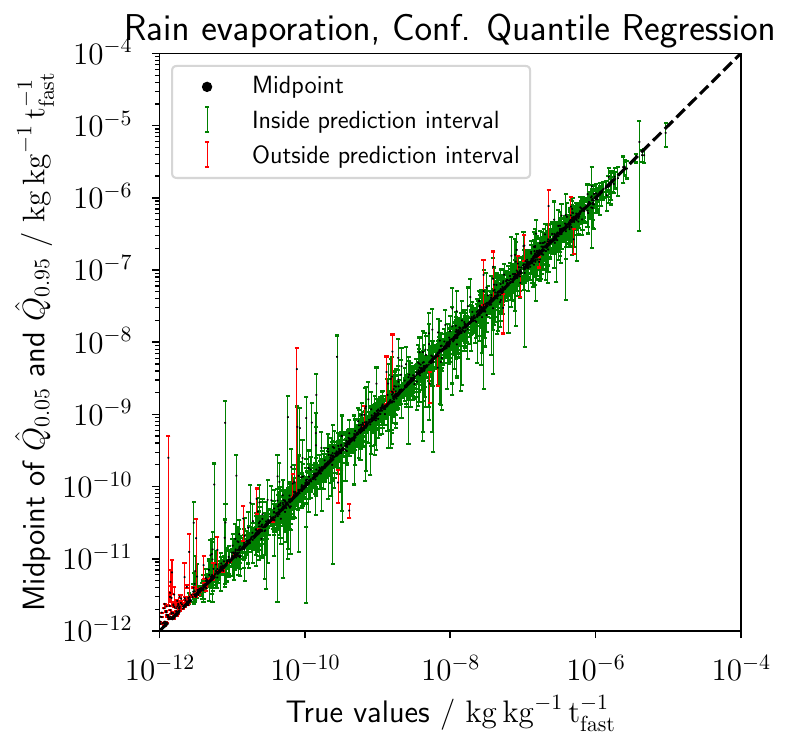} 
    \includegraphics[align=t, width=0.33\textwidth]{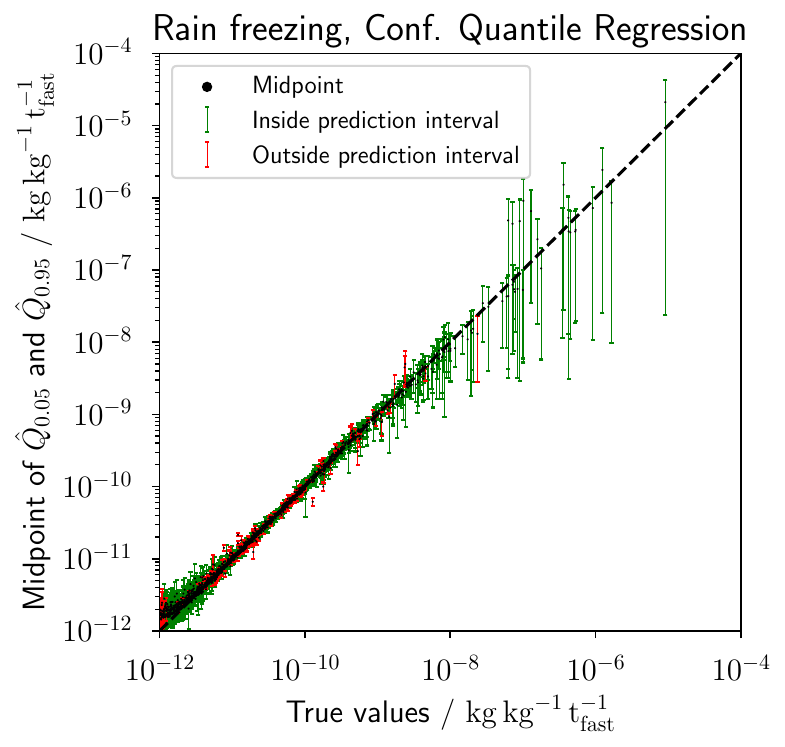}}
    \FIG{
    \includegraphics[width=0.33\textwidth]{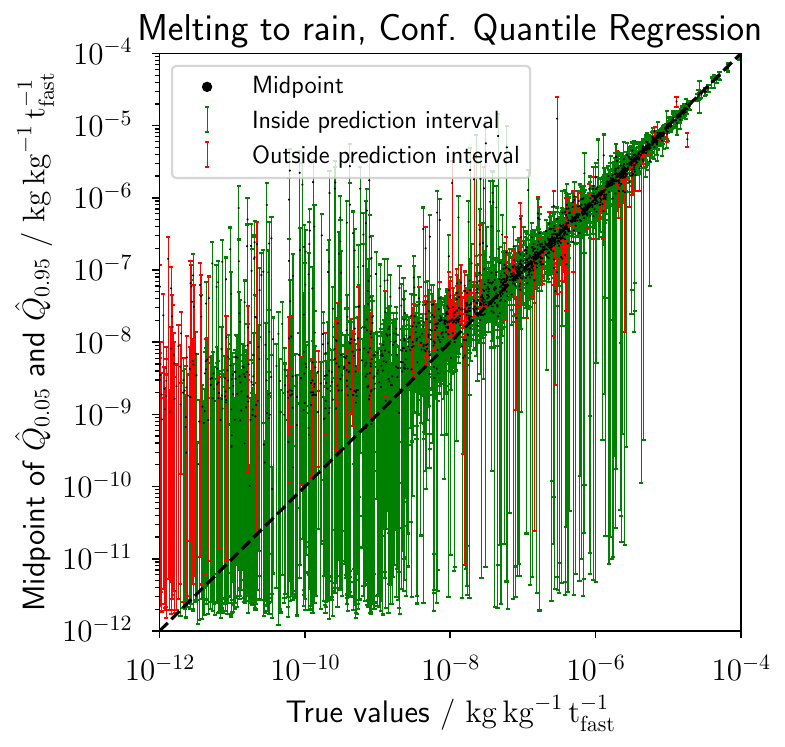}
    \includegraphics[width=0.33\textwidth]{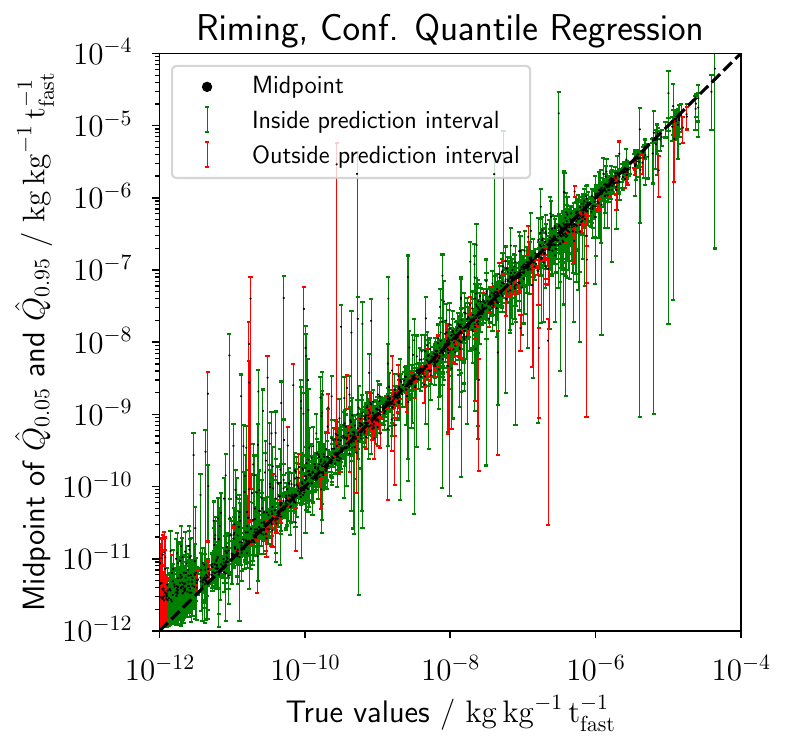} 
    }
    {\caption{Calibrated prediction intervals with conformalized quantile regression (CQR) for the accretion (QNN), rain evaporation (QRF), rain freezing (QXGB), melting to rain (QNN) and the total riming (QXGB) rate, obtained with the quantile regression model that yields the best PICP (in brackets, see Table~\ref{table:results_picp} in the main text). For better visualization, we only show 1500 randomly selected samples}\label{fig:intervals_cqr_app}}
\end{figure}

\subsection{Prediction interval coverage probability (PICP) and normalized mean prediction interval width (NMPIW)}\label{sec:picp_nmpiw_app}
In Figure~\ref{fig:scores_scp_app} and Figure~\ref{fig:scores_cqr_app}, we visualize the prediction interval coverage probability (PICP) and the normalized mean prediction interval width (NMPIW) for SCP and CQR for the accretion, rain evaporation, rain freezing, melting to rain and the total riming rate.
\begin{figure}[ht]
    \centering
    \FIG{\includegraphics[align=t, width=0.33\textwidth]{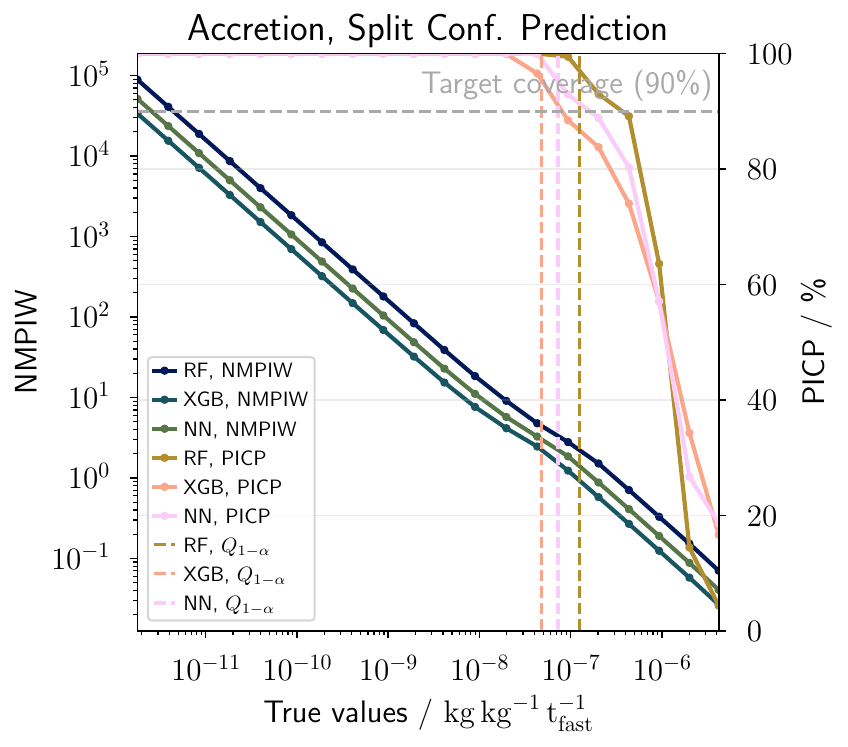}
    \includegraphics[align=t, width=0.33\textwidth]{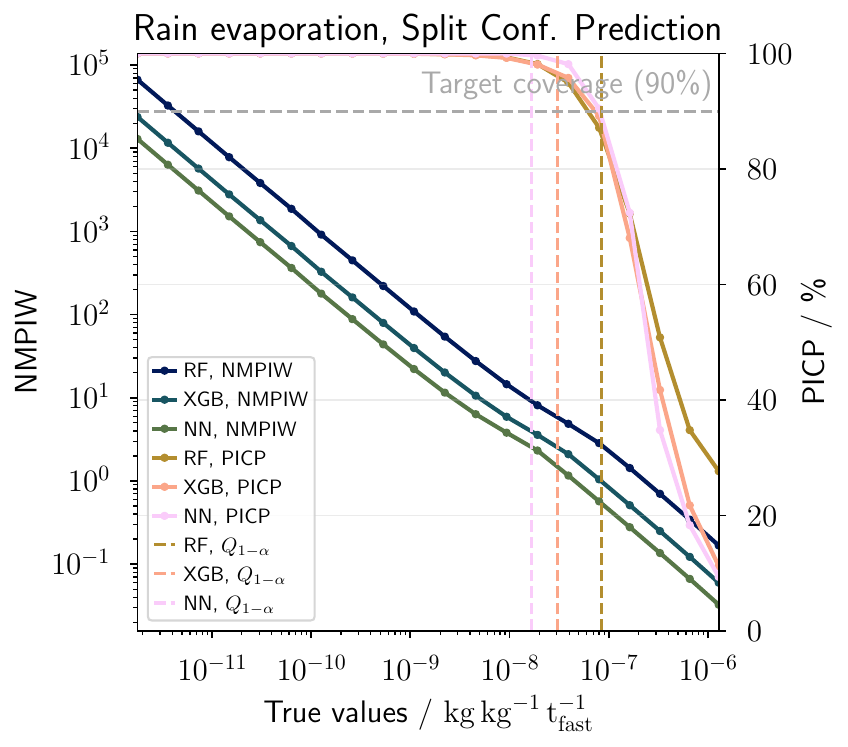} 
    \includegraphics[align=t, width=0.33\textwidth]{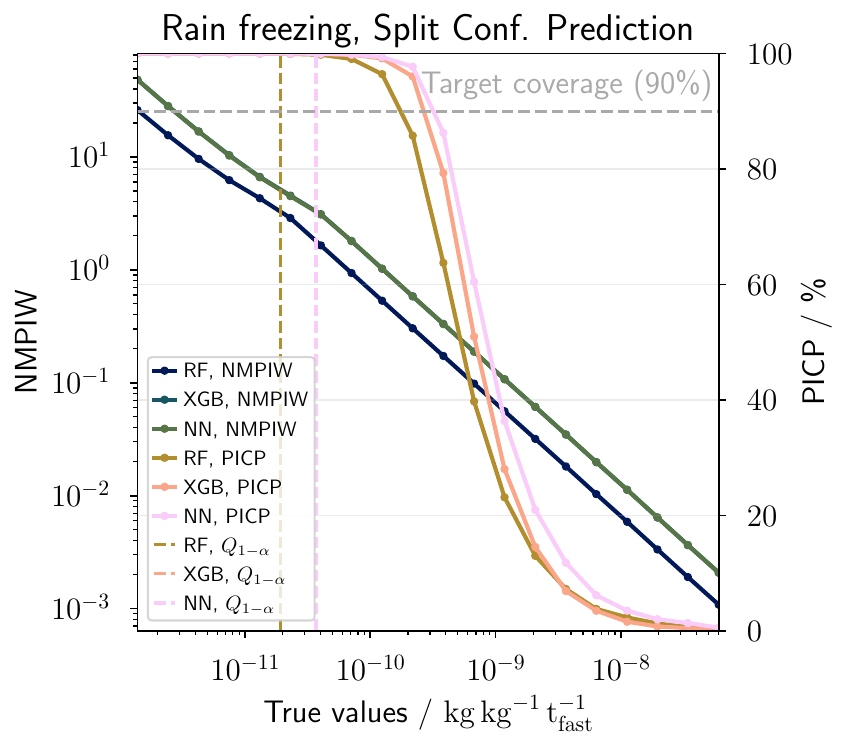}}
    \FIG{
    \includegraphics[width=0.33\textwidth]{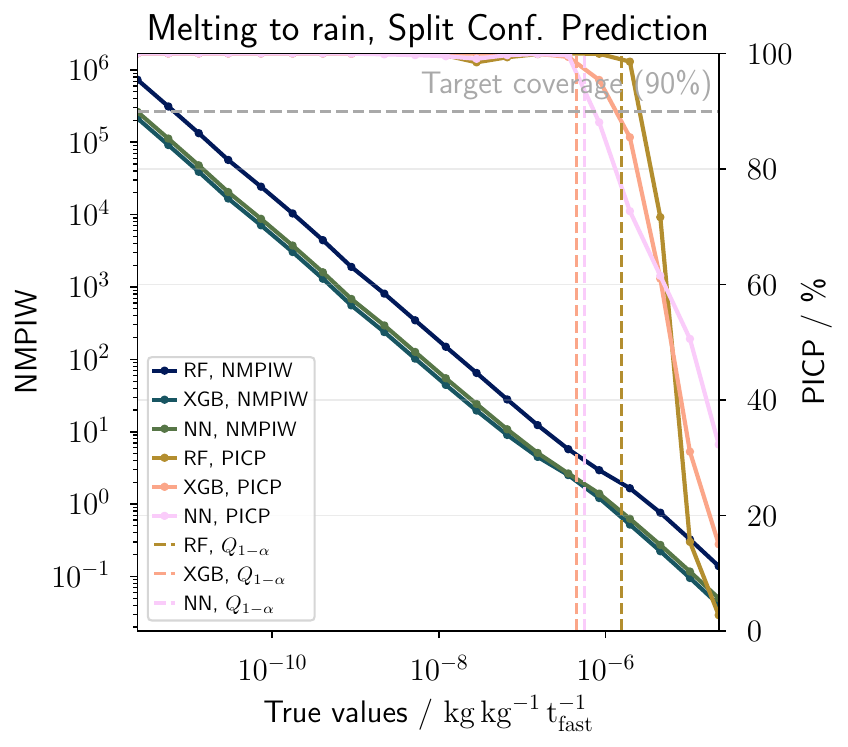}
    \includegraphics[width=0.33\textwidth]{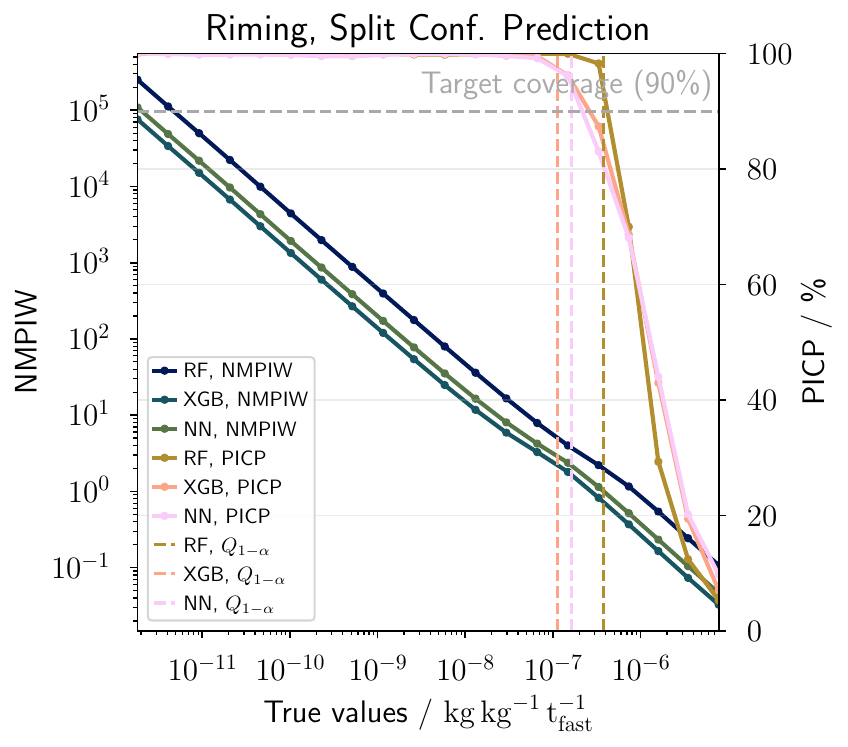} 
    }
    {\caption{PICP and NMPIW binned
by size of the true values for the accretion, rain evaporation, rain freezing, melting to rain and the total riming rate with split conformal prediction}\label{fig:scores_scp_app}}
\end{figure}
\begin{figure}[ht]
    \centering
    \FIG{\includegraphics[align=t, width=0.33\textwidth]{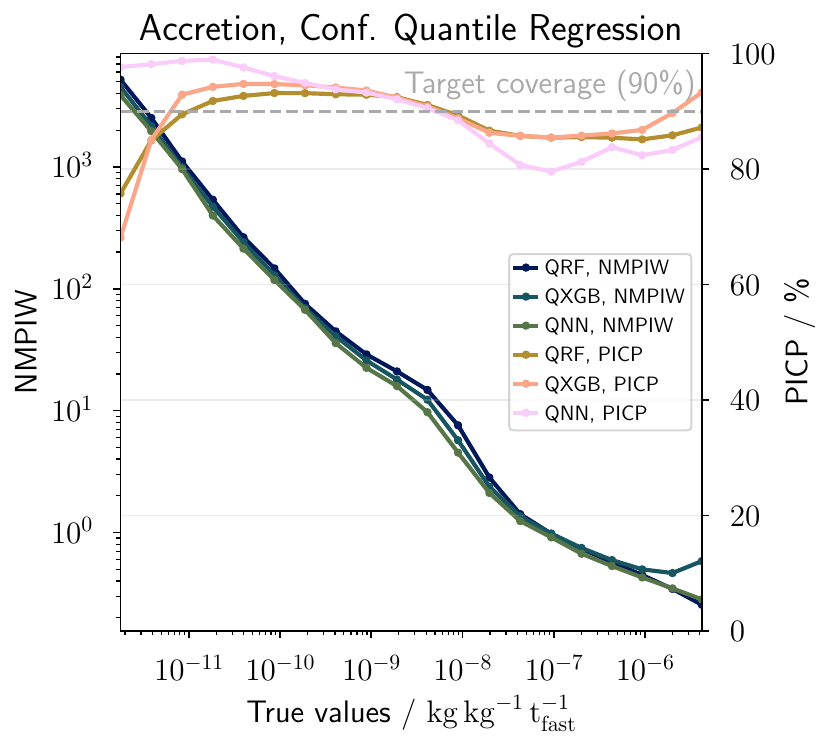}
    \includegraphics[align=t, width=0.33\textwidth]{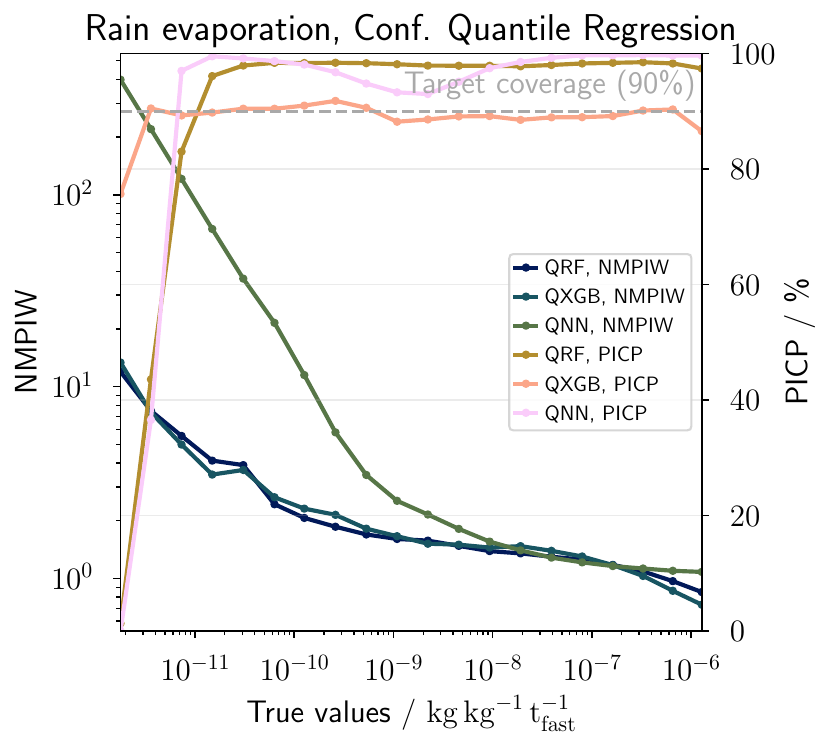} 
    \includegraphics[align=t, width=0.33\textwidth]{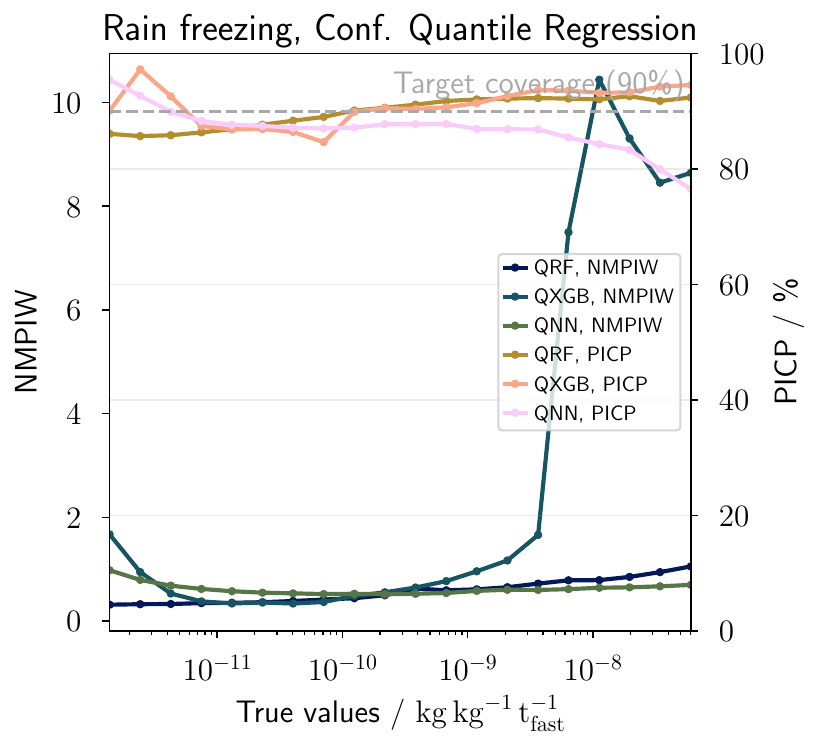}}
    \FIG{
    \includegraphics[width=0.33\textwidth]{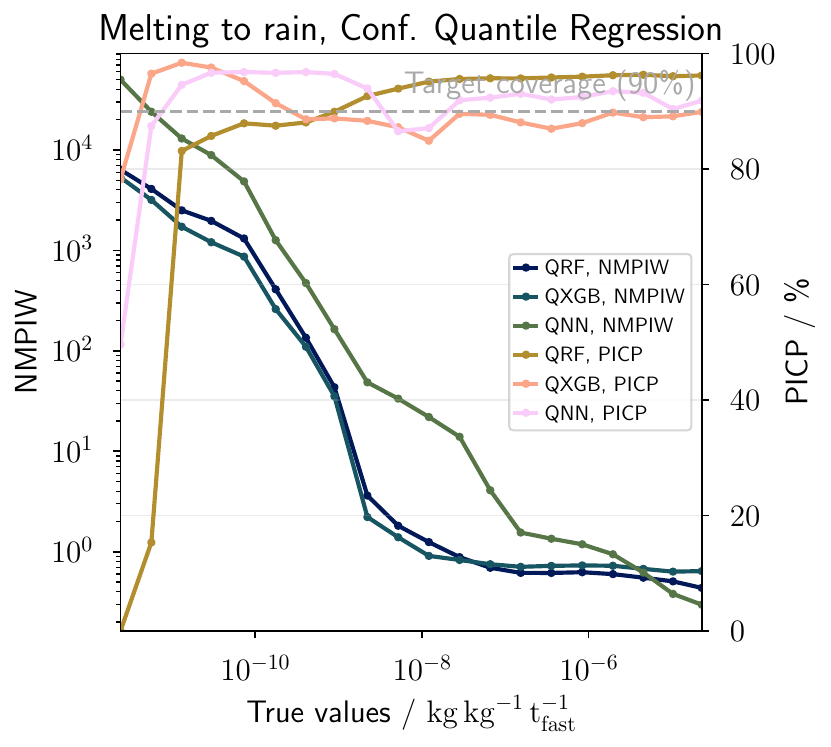}
    \includegraphics[width=0.33\textwidth]{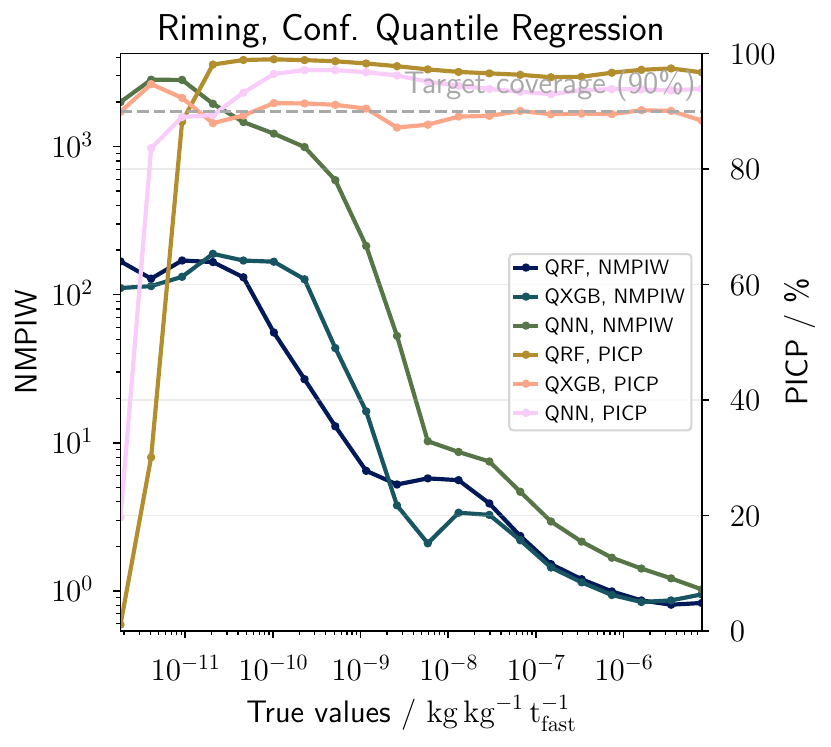} 
    }
    {\caption{PICP and NMPIW binned
by size of the true values for the accretion, rain evaporation, rain freezing, melting to rain and the total riming rate with conformalized quantile regression}\label{fig:scores_cqr_app}}
\end{figure}

\subsection{Heteroscedasticity analysis with absolute residuals}\label{sec:heteroscedasticity}
To assess heteroscedasticity, we consider the spread of the normalized residuals $\vert y_\text{true} - y_\text{pred}\vert /\vert y_\text{pred}\vert$ of the model predictions, which are a measure of how the spread of residuals varies with the range of true values. For the QR models, we again use the median of the uncalibrated predicted upper and lower quantile as $y_\text{pred}$. In Figure~\ref{fig:residuals_scp} and Figure~\ref{fig:residuals_cqr}, we show the interquartile range, $\text{IQR} = Q_{0.75}-Q_{0.25}$, and the $90\%$ spread, $Q_{0.95} - Q_{0.05}$, of the absolute residuals for all six process rates. The true values are binned by size. It is apparent that spread of the residuals is not constant but decreases with increasing size of the target value. This is for all process rates except for rain freezing, where the spread is relatively constant at low values and increases towards high true values, which is also the case for the total riming rate and SCP. Furthermore, in most cases with CQR, the 90\% spread is more than a magnitude larger than the IQR spread, indicating that the spread is large both in the bulk and in the tails of the distribution of absolute residuals. This indicates heteroscedasticity and further substantiates the superior performance of CQR in our case.
\begin{figure}[ht]
    \centering
    \FIG{\includegraphics[align=t, width=0.32\textwidth]{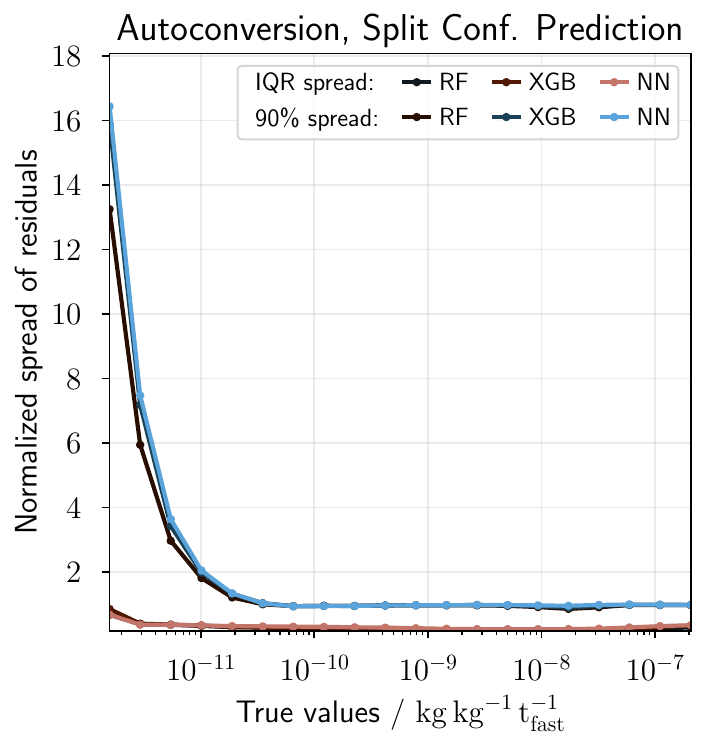}
    \includegraphics[align=t, width=0.33\textwidth]{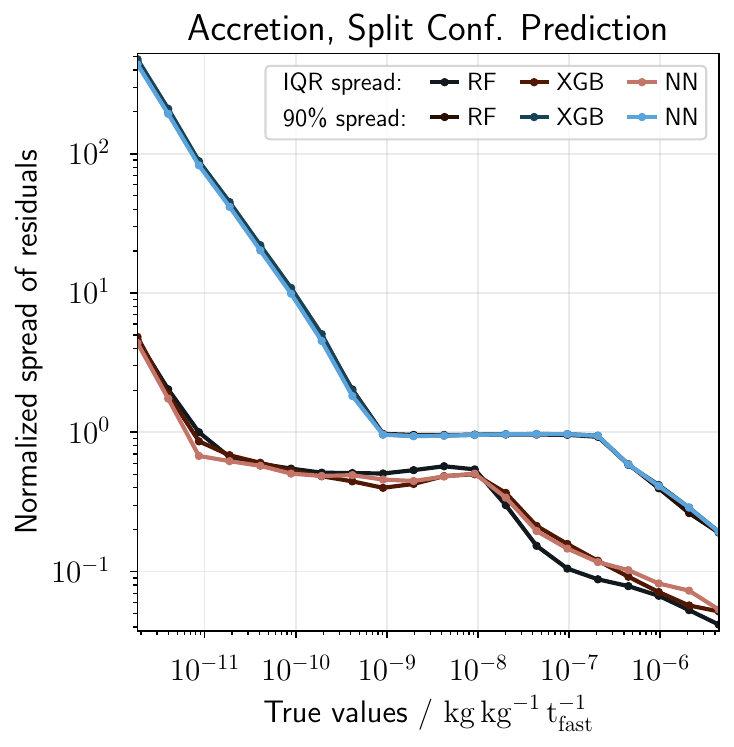}
    \includegraphics[align=t, width=0.33\textwidth]{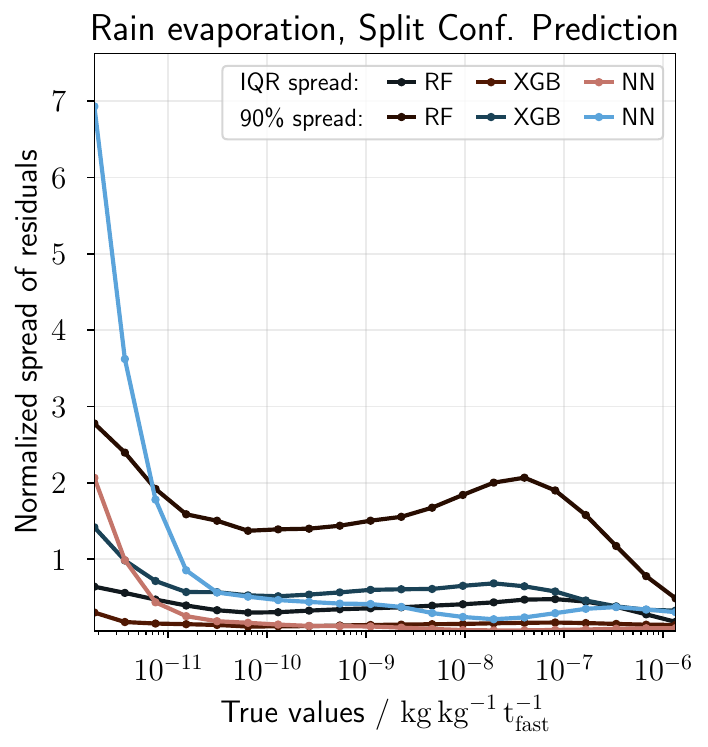} 
    }
    \FIG{\includegraphics[align=t, width=0.33\textwidth]{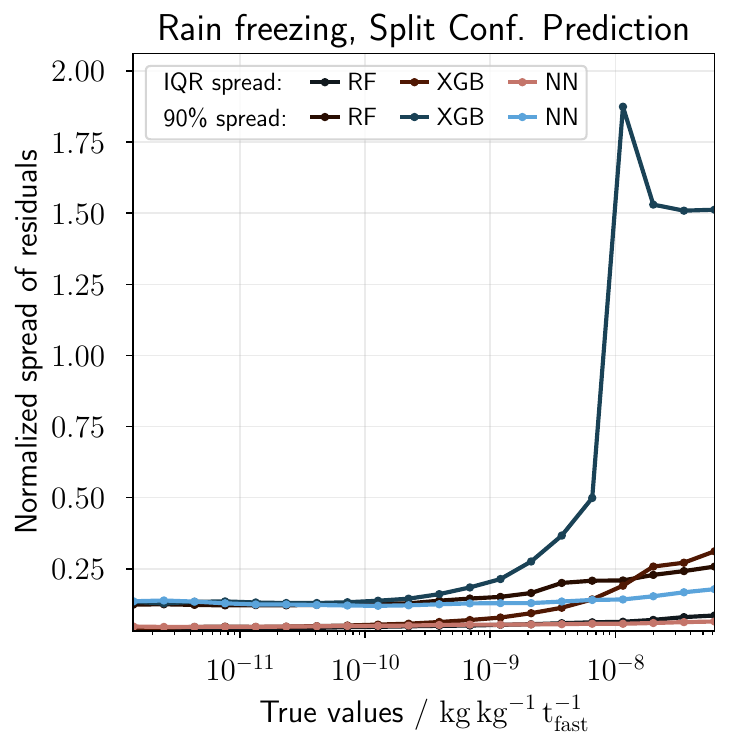}
    \includegraphics[align=t, width=0.33\textwidth]{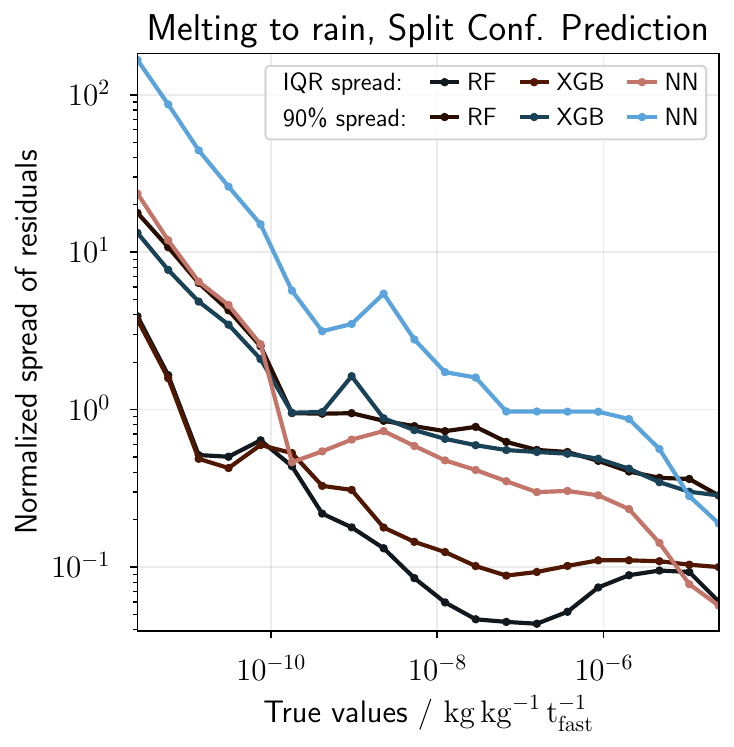}
    \includegraphics[align=t, width=0.33\textwidth]{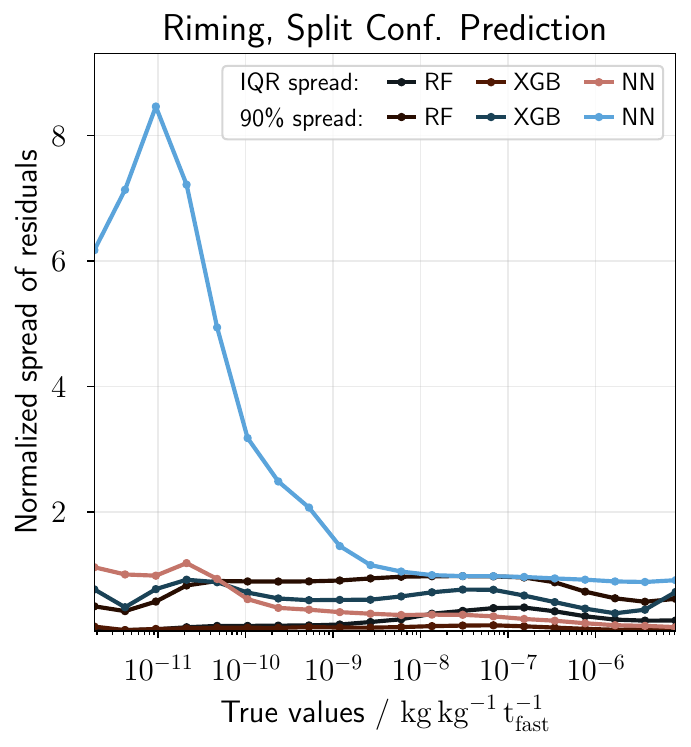} 
    }
    {\caption{Spread of absolute residuals based on the IQR and 90\% spread binned size of the true values for the autoconversion, accretion, rain evaporation, rain freezing, melting to rain and the total riming rate with split conformal prediction}\label{fig:residuals_scp}}
\end{figure}
\begin{figure}[ht]
    \centering
    \FIG{\includegraphics[align=t, width=0.32\textwidth]{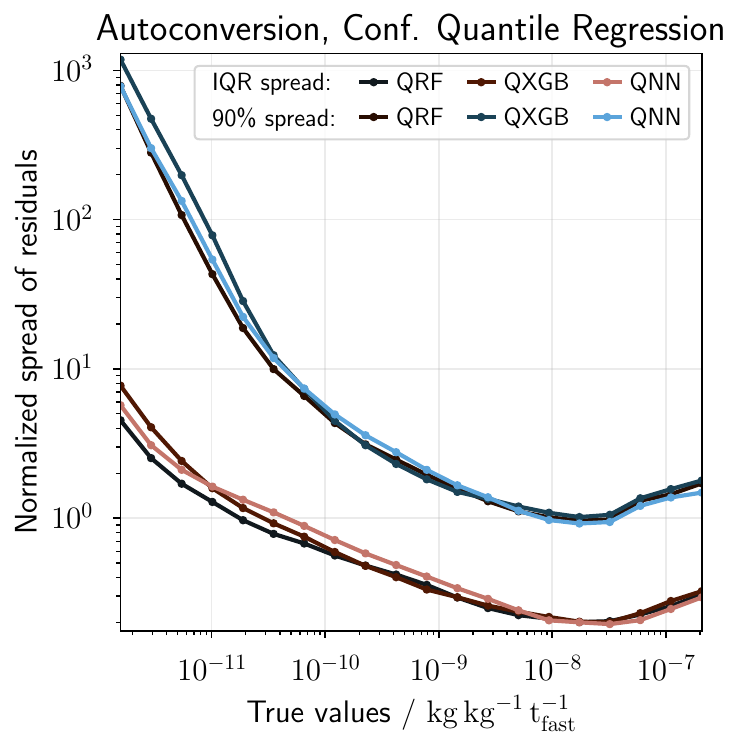}
    \includegraphics[align=t, width=0.33\textwidth]{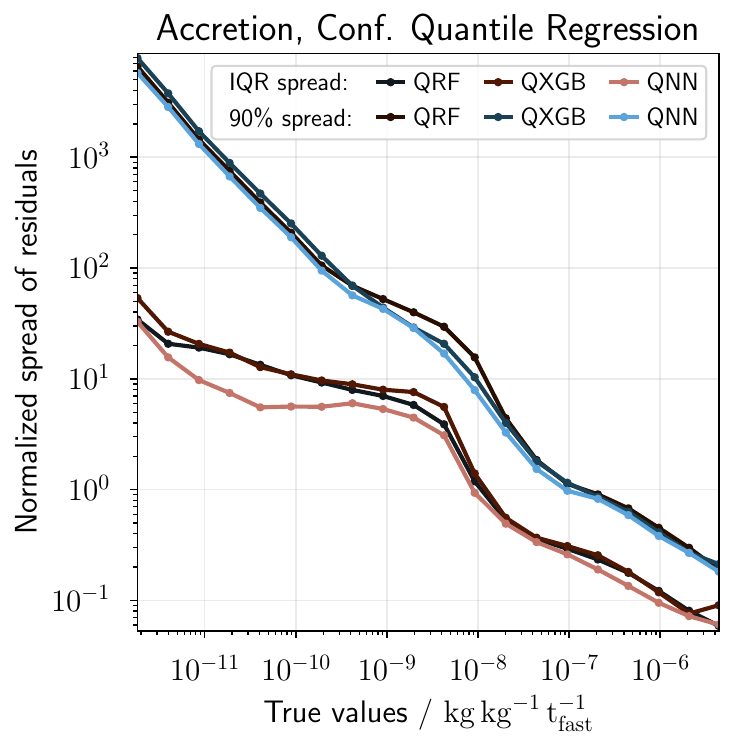}
    \includegraphics[align=t, width=0.33\textwidth]{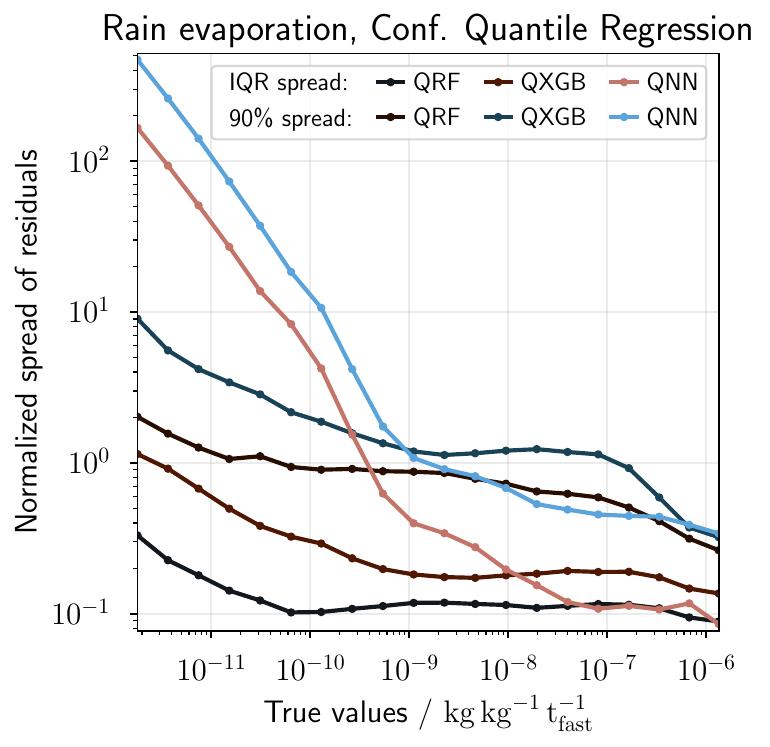}}
    \FIG{\includegraphics[align=t, width=0.33\textwidth]{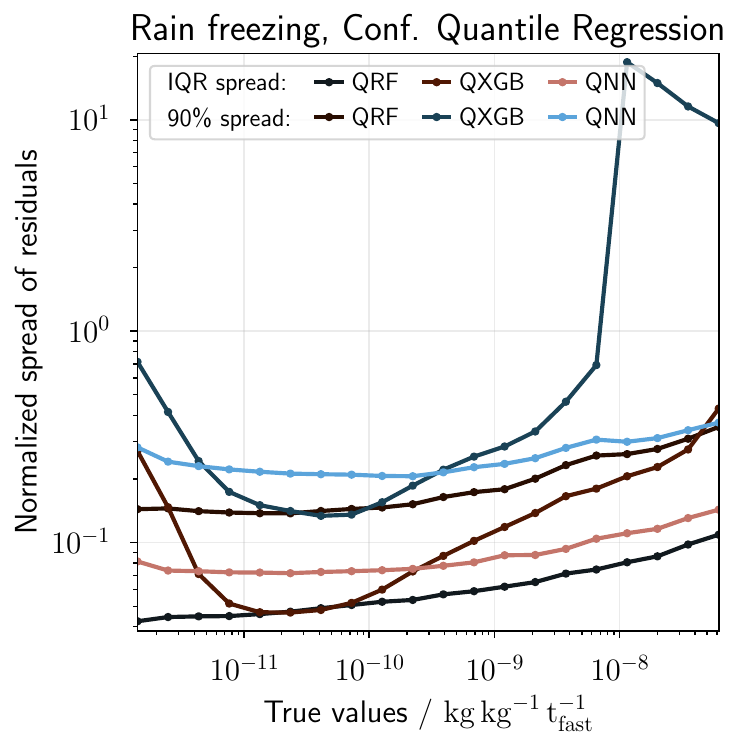}
    \includegraphics[align=t, width=0.33\textwidth]{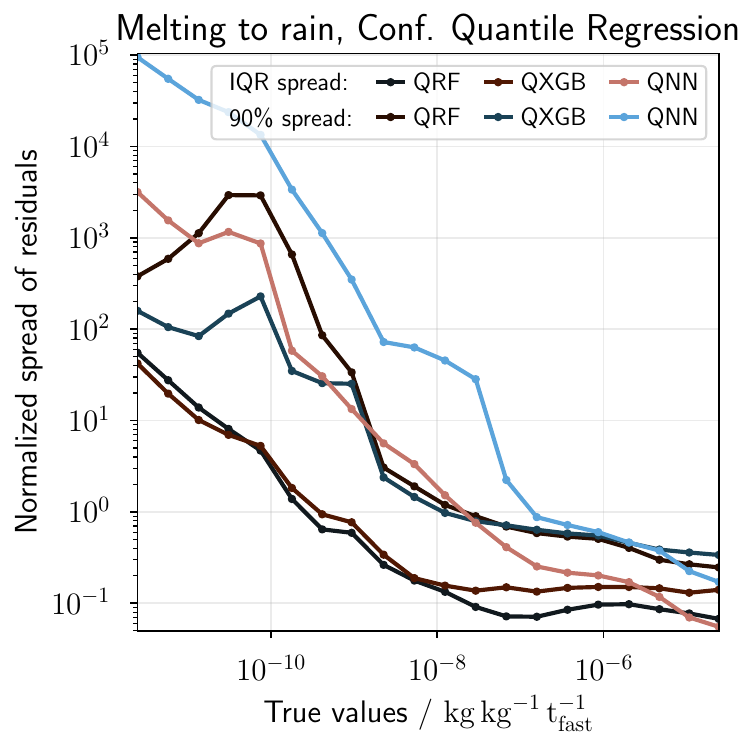}
    \includegraphics[align=t, width=0.33\textwidth]{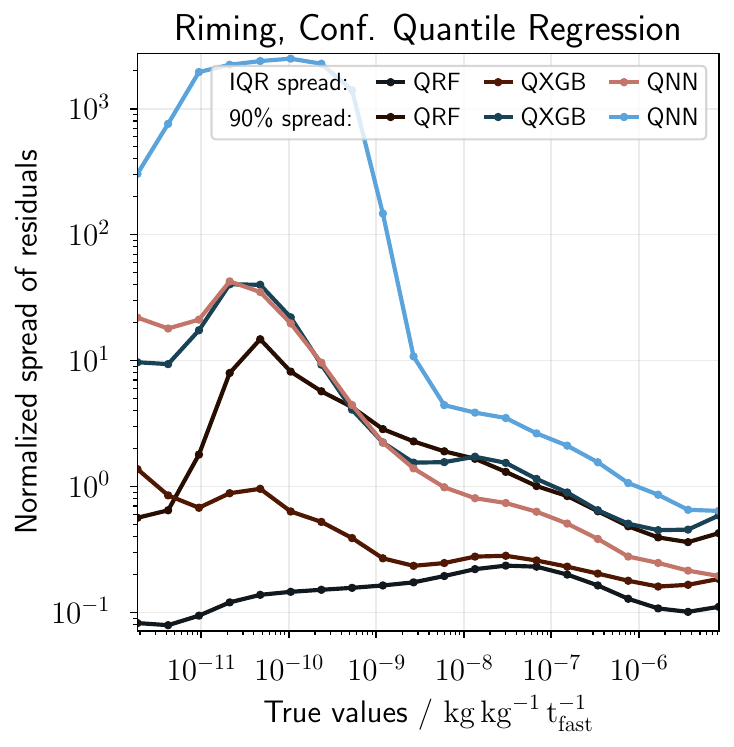} 
    }
    {\caption{Spread of absolute residuals based on the IQR and 90\% spread binned size of the true values for the autoconversion, accretion, rain evaporation, rain freezing, melting to rain and the total riming rate with conformalized quantile regression}\label{fig:residuals_cqr}}
\end{figure}
\end{appendix}
\end{Backmatter}

\end{document}